\documentclass{aastex61}
\shorttitle{HST Imaging of Bulge Globular Clusters: Observations, Data Reduction \& CMDs}
\shortauthors{R.E.~Cohen et al.}
\begin{document}
\title{Deep Hubble Space Telescope Imaging of Globular Clusters Towards the Galactic Bulge: Observations, Data Reduction, and Color-Magnitude Diagrams\footnote{Based on observations made with the NASA/ESA Hubble Space Telescope, obtained at the Space Telescope Science Institute, which is operated by the Association of Universities for Research in Astronomy, Inc., under NASA contract NAS 5-26555.  These observations are associated with program GO-14074}}

\correspondingauthor{Roger E. Cohen}
\email{rcohen@stsci.edu}

\author{Roger E. Cohen}
\affiliation{Space Telescope Science Institute, 3700 San Martin Drive, Baltimore, MD 21218, USA}
\affiliation{Departamento de Astronom\'{i}a, Universidad de Concepci\'{o}n, Casilla 160-C, Concepci\'{o}n, Chile}

\author{Francesco Mauro}
\affiliation{Departamento de Astronom\'{i}a, Universidad de Concepci\'{o}n, Casilla 160-C, Concepci\'{o}n, Chile}
\affiliation{Millenium Institute of Astrophysics, Av. Vicu\~na Mackenna 4860, 7820436 Macul, Santiago, Chile}
\affiliation{Instituto de Astronom\'{i}a, Universidad Cat\'{o}lica del Norte, Av. Angamos 0610, Casilla 1280, Antofagasta, Chile}

\author{Javier Alonso-Garc\'{i}a}
\affiliation{Unidad de Astronom\'{i}a, Fac. Cs. B\'asicas, Universidad de Antofagasta, Avda. U. de Antofagasta 02800, Antofagasta, Chile}
\affiliation{Millenium Institute of Astrophysics, Av. Vicu\~na Mackenna 4860, 7820436 Macul, Santiago, Chile}

\author{Maren Hempel}
\affiliation{Instituto de Astrof\'isica, Facultad de F\'isica, Pontificia
 Universidad Cat\'olica de Chile, Av.~Vicu\~na Mackenna 4860, 782-0436 Macul,
 Santiago, Chile}

\author{Ata Sarajedini}
\affiliation{Dept. of Astronomy, University of Florida, 211 Bryant Space Sciences Center, Gainesville, FL 32611, USA}
\affiliation{Department of Physics, Florida Atlantic University, 777 Glades Road, Boca Raton, FL 33431, USA}

\author{Antonio J. Ordo\~nez}
\affiliation{Insight Data Science Fellow}

\author{Douglas Geisler}
\affiliation{Departamento de Astronom\'{i}a, Universidad de Concepci\'{o}n, Casilla 160-C, Concepci\'{o}n, Chile}

\author{Jason S. Kalirai}
\affiliation{Space Telescope Science Institute, 3700 San Martin Drive, Baltimore, MD 21218, USA}

\begin{abstract}

The Galactic globular clusters (GGCs) located towards the Galactic bulge have generally been 
excluded from large-scale photometric GGC surveys due to severe total and differential extinction.  
Here, we present an overview of a \textit{Hubble Space Telescope} (HST) program designed to obtain deep, high spatial resolution multiband imaging of 16 poorly studied GGCs located towards the inner Galactic bulge and disk.  
In this first paper of a series resulting from these observations, we give an overview of 
target cluster selection, observations and data reduction procedures for optimizing the
resulting photometric catalogs.  Artificial star tests are used to compare the respective advantages of different data reduction strategies in terms of photometric and astrometric precision and photometric incompleteness.  We present
the resulting color-magnitude diagrams (CMDs) of all target clusters in several color-magnitude planes, along with CMDs of comparison fields from parallel observations.  For each target cluster, we summarize existing studies, and discuss their CMDs qualitatively in the context of these results.

\end{abstract}


\section{Introduction} \label{sec:intro}
\subsection{The Power of a Self-consistent Globular Cluster Sample}

Galactic globular clusters (GGCs) carry the imprint of the early formation
history of the Milky Way.
Therefore, by studying GGCs as an ensemble as well as individually, they serve as powerful
probes of Galactic dynamical and chemical evolution.
Deep space-based
photometric surveys have the power to yield
heretofore unprecedented results in these arenas, as exemplified by the \textit{HST} snapshot
survey performed with the Wide Field Planetary Camera 2 (WFPC2) \citep{piottosnap}, the Advanced Camera for Surveys (ACS) GGC Treasury Survey \citep{ataggc}, and the 
Ultraviolet Legacy Survey of GGCs \citep{piottolegacy}.  The scientific output 
facilitated by such massive, self-consistent datasets is difficult to overstate, so as an example we focus on the ACS GGC Treasury Survey:
By providing homogenous, deep photometry (SNR$\sim$10 at $\sim$0.2$M_{\odot}$)
for 65 GGCs, this survey has yielded a plethora of self-consistent results
covering a huge breadth
of topics.  These include
cluster distances and reddenings \citep{d10,cohensxphe}, both relative \citep{mf09} and absolute ages \citep{d10,v13},
luminosity and mass functions \citep{paustmf},
structural parameters and mass segregation \citep{goldsbury10,goldsbury13} and 
binary fractions \citep{milonebinfrac}.  The impact of these studies is due largely to the self-consistency of the observation and data reduction strategy, allowing direct comparisons across the largest sample to date.  However, such datasets inevitably have the added benefit of allowing in-depth analysis of individual cases of particular interest \citep[e.g.][]{milone1851,acssgr}.  Furthermore, by obtaining high-quality imaging of many previously unstudied targets, their legacy value may be exploited in complimentary future campaigns which yield both a time baseline for dynamical analyses \citep[e.g.][]{omegacenpm,samrapm,hstpromo1,hstpromo2,hstpromo3} as well as a broadened color baseline via imaging in complementary bandpasses \citep[e.g.][]{dotter6752,milone2808,gems2808,bssuv,bayesianrachel}

GGCs in the direction of the
Galactic bulge have been
almost entirely omitted from such large scale
studies due to severe total and differential extinction.
For example, 24 of the 26 GGCs within 2 kpc of the Galactic center according to the 
catalog of Harris (1996, 2010 edition, hereafter \citealt{h96}) lack a self-consistent age estimate, and nearly one third of them (8/26) lack any spectroscopic metallicity value from observations of individual stars\footnote{This fraction drops to 6/26 if one considers the benchmark integrated light studies by \citet{ZW84} and \citet{AZ88}, which included NGC 6333=M9 and Terzan 6 respectively.}.  
Here, we present a set of observations which build on existing Treasury programs, 
allowing the GGCs located towards the
bulge to be studied with the same level of scrutiny as the remainder of the GGC population.  

\subsection{The Globular Clusters Towards the Galactic Bulge}

There are already tantalizing hints that the GGCs of the bulge
hold valuable keys to Galactic astrophysics not found elsewhere.  This is exemplified by the complex stellar system Terzan 5, which was not recognized as being significantly outside the traditional definition of Milky Way GGCs until the discovery of a double horizontal branch (HB) by \citet{ter5hb}.  This discovery was bolstered by subsequent studies revealing that Terzan 5 is one of the most massive GGCs \citep{ter5sb}, and hosts stellar populations with a trimodal metallicity distribution spanning $\sim$1 dex in $[\rm{Fe/H}]$ \citep{ter5feh1,ter5feh2} and $\sim$7 Gyr in age \citep{ter5age}.  Shortly thereafter, PSF photometry of $ZYJHK_{S}$ imaging from the 
\textit{Vista Variables in the Via Lactea} survey \citep[VVV;][]{minnitivvv} revealed evidence of similar, albeit less well separated, double HBs
in two bulge GGCs which are also fairly massive and metal-rich, NGC 6440 and NGC 6569 \citep{maurohb}.  However, in contrast to Terzan 5, a metallicity spread has been ruled out as a contributor to the HB morphology of these two clusters \citep{cesar6440,6569spec_johnson} and in the case of NGC 6569, pulsational properties of its RR Lyrae variables also argue against helium enhancement \citep{kunder6569}.  More generally, the GGCs towards the Galactic center occupy unique areas of parameter space with respect to several key cluster parameters, including metallicity, HB morphology and concentration.  For example, of the GGCs with $[\rm{Fe/H}]$$<$-0.5, the majority are projected on the Galactic disk or bulge, including the only two (NGC 6528 and NGC 6553) that have approximately solar $[M/H]$ (e.g.~\citealt{maurocat}, \citealt{brunofors2}, \citealt{baitian6553}).  Furthermore, the incidence of candidate core-collapsed GGCs appears to increase substantially close to the Galactic center, and 17/28 of them have $R_{GC}$$\leq$3 kpc according to the \citet{h96} catalog.

Much of what we currently know about bulge GGCs is due to systematic efforts to characterize these clusters photometrically despite their higher total extinction and field star densities.  Bulge GGCs have been specifically targeted using ground-based imaging at optical (e.g.~\citealt{djorg2discov,ortolaniter1,bh261discov}, see \citealt{bicarev} for a review) and near-infrared wavelengths (\citealt{v10} and references therein\footnote{\url{http://www.bo.astro.it/~GC/ir_archive/Tab1_new.html}}) as well as a handful of clusters imaged with the first generation of space-borne near-IR arrays \citep{nicmos1,nicmos2}.  More recently, wide-field multi-wavelength photometric surveys such as the DECam Plane Survey \citep{decaps} and VVV are providing additional information on clusters falling within their survey areas \citep[e.g.][]{cohenvvv}.  However, with a few recent exceptions \citep[e.g.][]{cohen6544,6528hst,sara6624,pm6522} existing imaging generally lacks adequate depth and/or spatial resolution to probe below the main sequence turnoff (MSTO) of bulge GGCs.  Rather, luminous, evolved cluster red giant branch (RGB) and HB stars are used to estimate cluster distances, reddenings and metallicities, often via direct comparison with optically well-studied calibrating clusters \citep[e.g.][]{v04obs,ferraromethod}.  However, the distance and metallicity scales of the calibrating clusters have since seen substantial improvement, at least in a relative sense (see e.g.~figs.~18-19 of \citealt{cohenispi}).  Therefore, we aim to simultaneously harness the improvement in the parameters of optically well-studied GGCs together with deep, space-based imaging of the target bulge GGCs.  By combining improved cluster photometric parameters with measurements of their ages and structural parameters, we gain two related angles of attack on fundamental questions of Galactic evolution.

One angle of attack is via the age-metallicity relation (AMR) of GGCs,
which yields
critical constraints on the assembly of the Milky Way:
Recent evidence implies that metal-rich GGCs formed in the Galactic disk,
whereas the more metal-poor halo GCs were accreted
from satellite galaxies \citep[e.g.][]{leaman13}.
However, the formation history of the
Galactic bulge and its GGCs remains an open question, as 24 of
the 26 GGCs within 2 kpc of the Galactic center
\citep{h96} have been
excluded from \textit{all} recent systematic studies of GGC relative \textit{or} 
absolute ages \citep{deangeli,mf09,d10,v13,rachelage,chaboyerage}.  Age measurements of GGCs depend critically on their chemical abundances, but 
spectroscopic metallicity 
determinations with a precision of $<$0.1 dex for bulge GGCs are well within the reach of current observational capabilities via either high resolution optical spectroscopy \citep[e.g.][]{barbuyhp1b,cesar6440} or near-IR single- \citep[e.g.][]{valentiter1} or multi-object spectrographs \citep{baitian6553,jose6522}.  In this context, bulge GGC ages are the missing
piece of the AMR puzzle, and the AMR of the innermost GGCs is a critical probe of bulge formation and evolution.

The second angle of attack makes use of structural and morphological parameters.  
The structural and morphological parameters of GGCs located towards the Galactic bulge
are still based largely on either surface brightness profiles obtained from integrated light \citep{trager,mvdm}, or star counts which rely on bright upper-RGB stars \citep{bbsb}.  Both suffer from stochastic biases due to the low frequency of luminous cluster members as well as the effects of mass segregation within the GGCs \citep{noyolasb,goldsbury13}.  These issues can be alleviated by the simultaneous use of stellar number densities in place of integrated light, together with wide field photometry, but thus far such investigations have targeted mostly halo GGCs \citep[e.g.][]{miocchisb} since extinction and high field star densities complicate analyses at low Galactic latitudes.  For GGCs located towards the Galactic bulge, where line-of-sight
extinction varies on spatial scales of arcseconds \citep[e.g.][]{ag12,cohen6544}, the use of exclusively near-IR imaging is ideal.  Particularly when wide-field near-IR imaging
extending beyond the cluster outskirts is available, as in the present case (\citealt{cohenvvv}, see Sect.~\ref{targsect}), 
the marriage of such a catalog with high-spatial resolution near-IR imaging 
of cluster cores is necessary to circumvent strong total and/or variable extinction.  In at least one case, application of this strategy has revealed structural parameters substantially different from previous values based on optical integrated light \citep{liller1}.

Forthcoming papers in this series will present in-depth quantitative analyses of the aforementioned issues, while here we present an overview of our imaging campaign and data reduction strategies.  The next section describes our target cluster selection and overall  observation strategy.  In Section 3, we describe our data processing, including PSF photometry, artificial star tests, and the resulting completeness limits.  In Section 4, we present CMDs for each of our target clusters in several color-magnitude planes, discussing their features 
in the context of previous literature studies for each cluster.  Our results are summarized in the final section.

\section{Observations} \label{obssect}

\subsection{Target Cluster Selection \label{targsect}}

\begin{figure}
\plotone{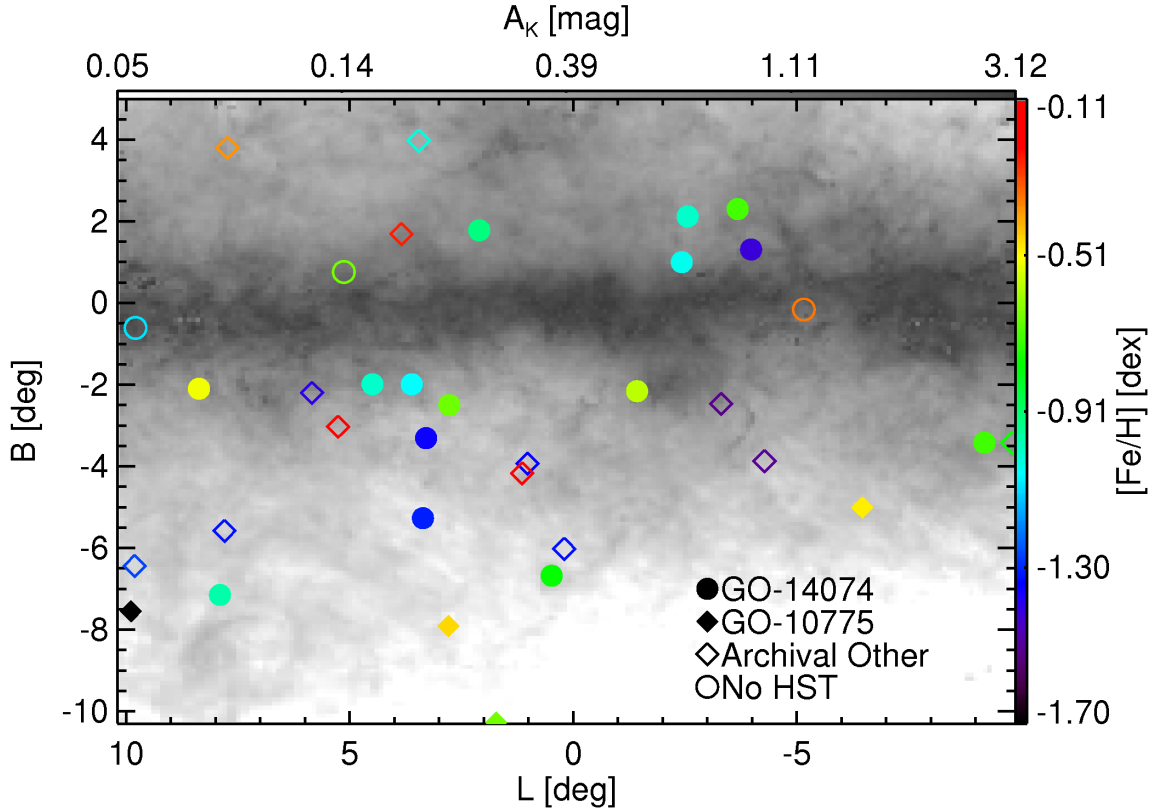}
\caption{Location of all GGCs from the \citet{h96} catalog located within the VVV bulge survey area.  Grey shading indicates the total $K_{S}$-band extinction from the map of \citet{gonzalezmap}, shown on a logarithmic greyscale (horizontal color bar).  Clusters are color-coded by \citet{h96} $[\rm{Fe/H}]$ (vertical color bar).  Our target clusters are shown as filled circles, clusters with deep ACS/WFC photometry from GO-10775 are shown as filled diamonds, clusters with archival \textit{HST} imaging from other programs are shown as open diamonds, and the three clusters with no \textit{HST} imaging are shown as open circles.  The target cluster FSR 1735, at $l$=-20.808, $b$=-1.854, is not shown. \label{lbmap}}
\end{figure}

Candidate target clusters were selected to meet several requirements: First, they have wide field near-infrared imaging from the VVV survey.  This imaging is crucial to the construction of radial number density profiles extending over the entire radial extent of each cluster.  Our catalogs of VVV PSF photometry carry the added benefit of providing a network of secondary astrometric standards much more numerous than either 2MASS or optical astrometric catalogs in heavily extincted regions.  Our targets are therefore spatially restricted to the VVV bulge (-10.0$\lesssim$$l$$\lesssim$10.5, -10.3$\lesssim$$b$$\lesssim$5.1) and disk (294.7$\lesssim$$l$$\lesssim$350.0, -2.25$\lesssim$$b$$\lesssim$2.25) survey areas.  In Fig.~\ref{lbmap}, we show a map of the VVV bulge survey area in Galactic coordinates, noting that only one of our target clusters (FSR 1735) lies in the VVV disk, rather than bulge, survey area\footnote{The reclassification of the cluster FSR 1716 as a candidate globular cluster, located in the VVV disk survey area, was recently announced by \citet{minnitinewgc}.}.  In Fig.~\ref{lbmap}, all GGCs have been color-coded by their metallicity from \citet{h96}, and all of our targets are shown as filled circles, while the four GGCs (NGC 6441, NGC 6624, NGC 6637=M69, NGC 6656=M22) with deep archival Treasury and Legacy survey imaging (which include the same ACS/WFC $F606W$ filter we employ) are shown as filled diamonds.

The second criterion imposed on our target list is that our target clusters are relatively poorly studied, and lack archival imaging of sufficient depth or spatial resolution to measure cluster ages (i.~e.~extending significantly faintward of the MSTO).  Our results therefore include the first unambiguous detection of the MSTO for these GGCs in many cases\footnote{For the more crowded and/or extincted targets, the MSTO is often difficult to discern even from VVV PSF photometry of cluster cores because of a combination of lower spatial resolution (0.339$\arcsec$/pix) and lower S/N relative to the color baseline than we obtain here.}.  However, it bears mention that for the handful of GGCs in the VVV survey area with deep archival $HST$ imaging not obtained as part of the Treasury Survey, shown as open diamonds in Fig.~\ref{lbmap}, age analyses are complicated by the heterogeneity of the available data.  Archival imaging of these clusters was obtained through different programs employing various filter and instrument combinations, but also different photometric depths, hampering the use of relative CMD indices for age measurements \citep[e.g.][and references therein]{v13}, particularly for clusters which saturate only slightly brightward of the MSTO.  

To arrive at the final target list, we decided to eliminate three additional clusters from our sample (2MASS GC02, Liller 1, UKS 1) which all have $E(B-V)>$3 according to the GGC catalog of \citet{h96}. 
This level of interstellar extinction is prohibitive to obtaining sufficiently deep WFC3/IR imaging in one orbit, and also to obtaining ACS/WFC optical imaging below the MSTO in a small number of orbits.   
For this reason, we have also excluded several additional candidate GGCs \citep{fsr1767,minniticl001,kron49} pending independent constraints from ground-based observations employing either spectrocopy \citep{gc02spec}, time series imaging of variable stars \citep{javiergcrrl}, and/or high spatial resolution near-IR follow-up imaging \citep[e.g.][]{liller1}.

Our target clusters are listed in Table \ref{targtab} along with their equatorial and Galactic coordinates and total exposure times for the primary field in each filter.

\subsection{Primary Fields}
\subsubsection{Observing Strategy}

Each primary field is observed for one orbit with ACS/WFC in $F606W$, and one orbit with WFC3/IR, splitting the orbit between $F110W$ and $F160W$.  Despite large total and differential extinction towards many of our target clusters, there are three reasons why the ACS/WFC imaging provides a valuable complement to WFC3/IR:

\begin{enumerate}
\item Spatial resolution: Since WFC3/IR has 0.13$\arcsec$/pix, photometry of cluster cores is crowding-limited.  The ability to leverage the increased spatial resolution of ACS/WFC (0.05$\arcsec$/pix) together with the WFC3/IR observations significantly enhances our ability to recover stars in the innermost regions of the target clusters, and improves overall positional accuracy (see Sect.~\ref{compreducsect}).

\item Field of view: The ACS/WFC field of view (202$\times$202$\arcsec$ per exposure) is large enough compared to the WFC3/IR field of view (123$\times$136$\arcsec$ per exposure) that the WFC3/IR observations of all target clusters are colocated completely within the ACS/WFC field of view.  This allows us to provide a complimentary bandpass of optical imaging for all relatively bright WFC3/IR sources. 
In addition, number density profiles covering the entire radial extent of target GGCs require a range of radial overlap between wide-field (typically shallower) imaging of the cluster outskirts and high-resolution imaging of the cluster cores, so the ACS/WFC observations give the option to choose increased radial coverage over increased photometric depth.  Moreover, given the higher spatial resolution of ACS/WFC over WFC3/IR, the larger field of view also provides larger radial coverage of each cluster for future dynamical investigations which could utilize our ACS/WFC imaging as a first epoch to measure relative proper motions.

\item Color baseline: The precision of CMD-based analyses, including differential reddening corrections and age measurements, depends on the ratio of the observed color range to the photometric measurement uncertainty at a given CMD location.  By complementing deep near-IR imaging with a bandpass of optical photometry, we can exploit smaller relative uncertainties in 
photometric measurements for the more optically luminous cluster members.
\end{enumerate}

\subsubsection{Filters}

Our choice of filters for each instrument was made not only to maximize throughput, but also to complement existing imaging.  This approach allows direct comparison to archival observations of both template GGCs as well as template bulge/disk fields in the native photometric system.
We choose $F606W$ for ACS/WFC imaging, since it allows a broad color baseline when paired with near-IR filters from either \textit{HST} or ground-based facilities.  At the same time, direct comparisons can be made with the large set of GGCs observed in GO-10775 \citep{ataggc} and additional archival imaging \citep[e.g.][]{dotteroh,6528hst} using, for example, $\Delta$$F606W$ as a distance-independent CMD-based metric.  Similarly, for WFC3/IR observations, $F110W$ and $F160W$ not only yield the best compromise between color baseline and reddening insensitivity, but also allow a direct comparison to archival imaging of bulge and disk fields and template GGCs spanning a range of metallicities 
\citep[e.g.][]{brownwfc3,holtziso,matteoir}.  

\subsubsection{Exposure Sequences}

The exposure sequence for each instrument was designed based on proven strategies from previous imaging campaigns to obtain the best compromise between photometric depth and dynamic range.  The ACS/WFC observations were obtained using the strategy from GO-10775, in which a single orbit is used to obtain one short exposure and four dithered long exposures in $F606W$.  The dither step is large enough to cover the chip gap and provide contiguous imaging over the entire field of view, and the short exposure times are varied between 10s and 60s based on both orbital visibility constraints and total optical extinction towards each cluster.  This allows the cluster HB to remain unsaturated in all cases, while obtaining maximum S/N within a single orbit for heavily extincted clusters with fainter apparent HB magnitudes in $F606W$.  

For WFC3/IR primary observations, we use a strategy similar to GO-11664, with the goal of obtaining the deepest possible photometry within one orbit without saturating the cluster HBs.  Therefore, we interlace $F110W$ and $F160W$ exposures using STEP100 sequences of varying length with NSAMP ranging from 5 to 12.  Observations are obtained at three dither points in both filters to mitigate cosmetic defects in the WFC3/IR array.  

\begin{deluxetable}{lcllcccccc}
\tablecaption{Target Clusters \label{targtab}}
\tablehead{
\colhead{Cluster} & \colhead{Other Name} & \colhead{RA (J2000)\tablenotemark{a}} & \colhead{Dec (J2000)\tablenotemark{a}} & \colhead{l\tablenotemark{a}} & \colhead{b\tablenotemark{a}} & \colhead{t$(F606W)$} & \colhead{t$(F110W)$} & \colhead{t$(F160W)$} \\ \colhead{} & \colhead{} & \colhead{hh:mm:ss.s} & \colhead{-dd:mm:ss.s} & \colhead{$^\circ$} & \colhead{$^\circ$} & s & s & s
}
\startdata
BH261 & ESO 456-78 & 18:14.1 & -28:37 & 3.37 & -5.27 & 1990 & 1246.154 & 1271.155 \\
Djorg 2 & ESO 456-38 & 18:01:49.1 & -27:49:33.0 & 2.7636 & -2.5085 & 2000 & 1246.154 & 1271.155 \\
FSR 1735 & 2MASS GC-03 & 16:52:12 & -47:03.3 & -20.808 & -1.854 & 2152 & 1296.155 & 1346.156 \\
HP 1 & ESO 455-11 & 17:31:05.2 & -29:58:54.0 & -2.5748 & 2.1150 & 2020 & 1246.154 & 1271.155 \\
NGC 6540 & Djorg 3 & 18:06:08.6 & -27:45:55.0 & 3.2851 & -3.3130 & 1990 & 1246.154 & 1271.155 \\
NGC 6569 & ESO 456-77 & 18:13:38.88 & -31:49:35.2 & 0.4814 & -6.6809 & 2024 & 1246.154 & 1271.155 \\
NGC 6638 & Gcl 95 & 18:30:56.25 & -25:29:47.1 & -7.8977 & -7.1530 & 1990 & 1246.154 & 1271.155 \\
Palomar 6 & ESO 520-21 & 17:43:42.2 & -26:13:21.0 & 2.0921 & 1.7801 & 2020 & 1246.154 & 1271.155 \\
Terzan 1 & ESO 455-23 & 17:35:47.8 & -30:28:11.0 & -2.4311 & 0.9958 & 2043 & 1246.154 & 1271.155 \\
Terzan 2 & ESO 454-29 & 17:27:33.2 & -30:48:07.8 & -3.6802 & 2.2978 & 2036 & 1246.154 & 1271.155 \\
Terzan 4 & Gcl 66.1 & 17:30:38.9 & -31:35:44.0 & -3.9762 & 1.3080 & 2036 & 1246.154 & 1271.155 \\
Terzan 6 & ESO 455-49 & 17:50:46.4 & -31:16:31.0 & -1.4286 & -2.1618 & 2036 & 1246.154 & 1271.155 \\
Terzan 9 & Gcl 80.1 & 18:01:38.8 & -26:50:23.0 & 3.6031 & -1.9888 & 2012 & 1246.154 & 1271.155 \\
Terzan 10 & ESO 521-16 & 18:02:57.4 & -26:04:00.0 & 4.4207 & -1.8629 & 2022 & 1246.154 & 1271.155 \\
Terzan 12 & ESO 522-1 & 18:12:15.8 & -22:44:31.0 & 8.3581 & -2.1008 & 2010 & 1246.154 & 1271.155 \\
Ton 2 & ESO 333-16 & 17:36:11 & -38:33.0 & 9.201 & -3.421 & 2020 & 1271.155 & 1271.155 \\
\enddata
\tablenotetext{a}{Literature values for cluster centers are taken from the SIMBAD database.}
\end{deluxetable}

\subsection{Parallel Fields}

\begin{figure}
\plotone{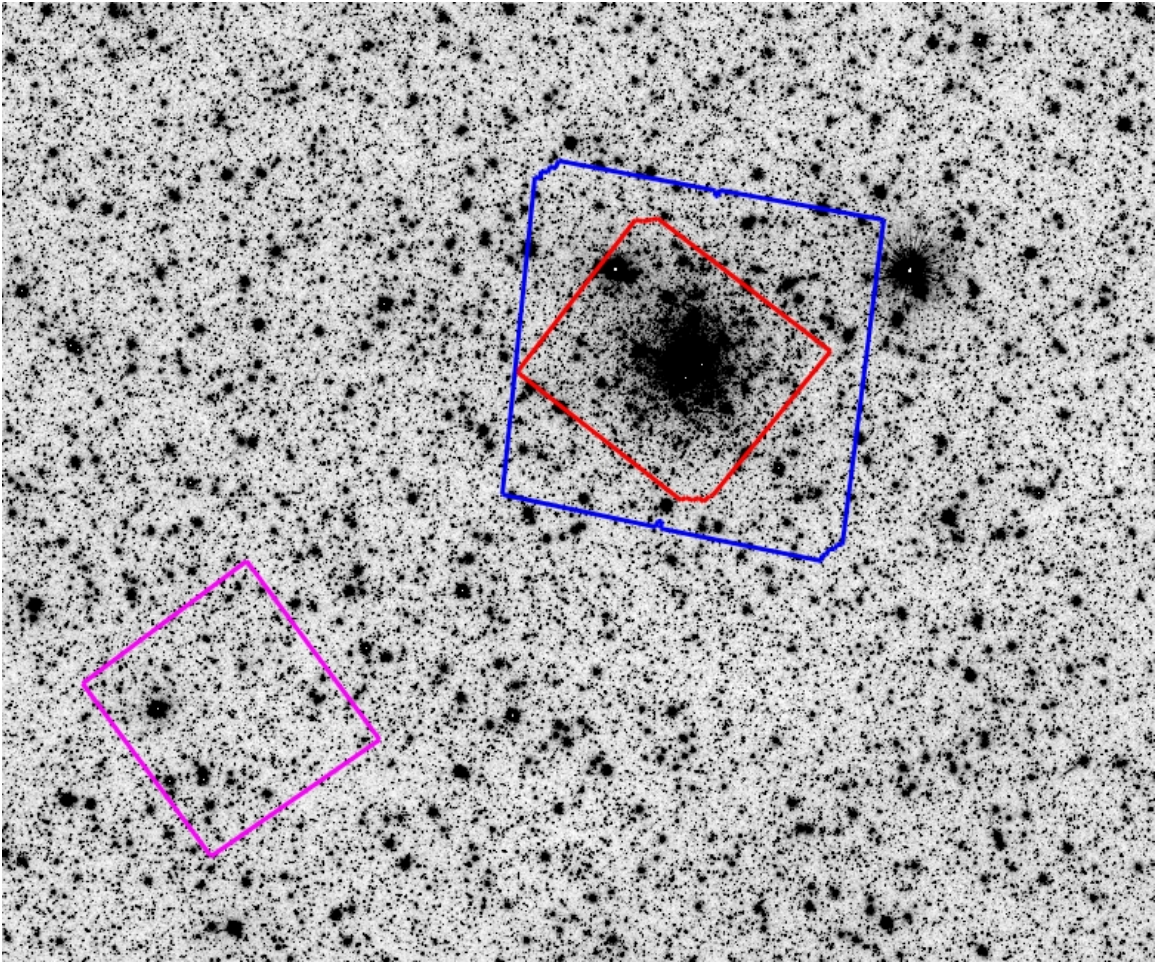}
\caption{A 12$\arcmin$$\times$10$\arcmin$ portion of a VVV $K_{S}$-band image illustrating the relative
orientations of the \textit{HST} GO-14074 fields for the example of Terzan 1.  The primary fields are shown for ACS/WFC (blue) and WFC3/IR (red), and the WFC3/IR parallel field is shown in magenta.  
North is up and east is to the left, and this VVV image has 0.339$\arcsec$/pix and was observed under seeing of $\sim$0.72$\arcsec$. \label{fovfig}}
\end{figure}

Imaging of the primary field of each target cluster is supplemented with WFC3/IR $F110W$ and $F160W$ imaging of a parallel field centered $\sim$6 arcmin from the center of the primary field.  As these
exposures are obtained during the dithered ACS/WFC observations of the primary field, the need to 
read out and dither ACS/WFC limited the available WFC3/IR observing sequences.  As a result, 
for each parallel field we employed one STEP100 NSAMP=9 and one STEP100 NSAMP=12 WFC3/IR sequence per filter, for a total per-filter exposure time of 898.463s.  In this way, we are able to attain a photometric depth comparable to the primary field WFC3/IR observations, noting that the shorter exposure times are counterbalanced to some extent by the reduced impact of crowding in the parallel fields far from the cluster centers.  We placed no requirement on the relative orientation between the primary and parallel fields to optimize schedulability of the observations, so the angle between primary and parallel fields varies from cluster to cluster.  However, to illustrate their relative sizes and orientations, the location of the parallel field relative to the primary fields is shown for the example case of Terzan 1 in Fig.~\ref{fovfig}. 

\section{Photometry}

We perform preprocessing and PSF photometry using the latest version of \texttt{Dolphot} 2.0 \citep{dolphin}.
\texttt{Dolphot} provides preprocessing and photometry modules tailored to each \textit{HST} camera, including pre-computed PSFs for each filter of each instrument and, crucially in our case, the ability to photometer images from multiple cameras simultaneously.  A description of \texttt{Dolphot}, as well as recommended photometry parameters for each instrument (hereafter referred to as "default" values), can be found in the manuals available at the \texttt{Dolphot} website\footnote{\url{http://americano.dolphinsim.com/dolphot/}}.

\subsection{Preprocessing and Image Alignment \label{preprocsect}}

The best multi-band ($F606W,F110W,F160W$) photometry resulted from a two-tier data reduction strategy. 
First, images from each camera (ACS/WFC or WFC3/IR) are preprocessed separately, largely according
to prescriptions in the \texttt{Dolphot} manual.  Briefly, pixels flagged in the data quality images
are masked from any further incorporation, and each science image is split into a single extension .fits
file per chip (one for WFC3/IR, one for each of the two ACS/WFC chips per image).  Next, the
\texttt{calcsky} task is used to create a sky image corresponding to each individual 
science image. 
Since we ultimately found that setting \texttt{FitSky}=2, in which the sky is fit inside the PSF radius but outside the photometry aperture, yielded the best photometry, 
we use the higher-resolution positive step size (\texttt{step}=2 for WFC3/IR or 4 for ACS/WFC) 
when generating the sky images.

After preprocessing, 
initial \texttt{Dolphot} photometry runs were performed on the ACS/WFC and WFC3/IR images separately.  This is due in part to offsets of as much as $\sim$1 arcsec in the world coordinate system (WCS)
information contained in the headers of the ACS/WFC versus WFC3/IR images, hampering simultaneous alignment of all images to a single reference frame.  Therefore, relatively bright ($F160W$$<$20) stars passing our photometric quality cuts (see below) were cross-matched between the two 
catalogs from the separate initial \texttt{Dolphot} runs to calculate a positional offset between the ACS/WFC and WFC3/IR distortion corrected reference frames.  This positional offset was measured using an iterative 5-sigma clip until convergence was indicated by a change of $<$1 mas in positional offset.  The final number of stars used for this calculation ranged from 2008 to 6851 across our target clusters (with a median of 4028), and rms deviation of the final offset among the stars used for matching had a median of 7.1 mas across our cluster sample, with a standard deviation of 1.9 mas.  The final positional shifts for each cluster were then applied to
the \texttt{CRVAL1} and \texttt{CRVAL2} header keywords of all of the WFC3/IR images to place them on the
same \textit{relative} astrometric reference frame as the ACS/WFC images.  At this point, a final photometry run was performed on all images simultaneously, using a deep drizzled, cosmic ray cleaned, distortion corrected ACS/WFC \texttt{.drc} image as a positional reference frame (brightness measurements are not performed on the drizzled reference image, rather they are calculated exclusively from the ensemble of individual \texttt{.flc} or \texttt{.flt} science exposures).  

\begin{figure}
\plotone{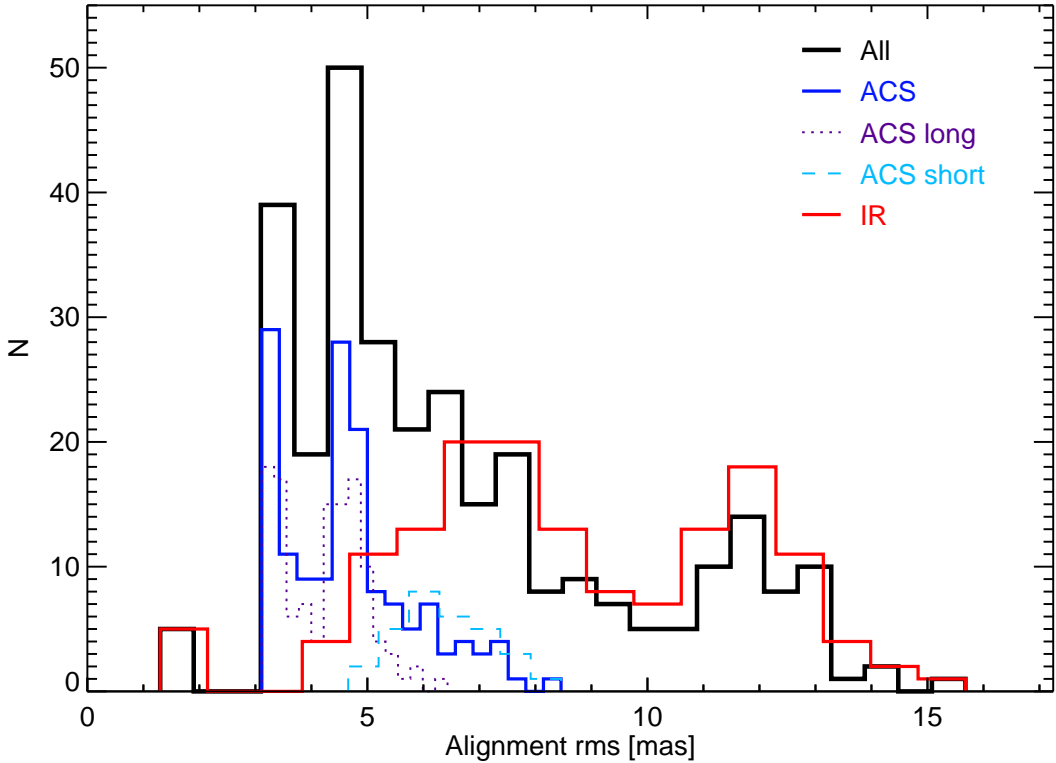}
\caption{Histograms of the \texttt{Dolphot} initial alignment rms over all images of all target clusters.  Different line thicknesses, styles and colors are used to indicate various subsamples (shown using different bin sizes for clarity) as indicated in the upper right-hand corner.}   
\label{alignfig} 
\end{figure}

In light of the difference in pixel scale and field of view between ACS/WFC and WFC3/IR, the improvement in photometric accuracy and completeness gained by leveraging multiple images together depends critically on the ability to accurately align all individial science images via a common
reference frame.  Therefore, we searched for the best set of \texttt{Dolphot} alignment parameters by varying
the values of \texttt{UseWCS}, \texttt{Align}, \texttt{AlignTol}, \texttt{AlignStep} and \texttt{AlignIter} in search of the smallest alignment rms.   The best final set of alignment parameters is given in Table \ref{doltab} for parameters where the optimal adopted value differed from the recommendation of the \texttt{Dolphot} manual.  The total number of stars used for alignment varied between 2667 and 50109 per cluster, and histograms of the alignment rms across our full sample of target clusters is shown in Fig.~\ref{alignfig}.  There, it is apparent that the ACS/WFC images typically yielded the smallest alignment rms relative to the reference frame due to their increased spatial resolution.  Furthermore, the long ACS/WFC exposures typically have a smaller alignment rms than the short exposures due to their increased photometric depth (i.e.~larger sample size).  Meanwhile, the WFC3/IR exposures yield a larger alignment rms in some cases due to their lower spatial resolution, and we found that among the IR exposures with an alignment rms $>$10 mas, 78$\pm$16\% were in the $F110W$ filter, whereas all $F160W$ images had an alignment rms $<$12.2 mas, contributing to the bimodality seen in Fig.~\ref{alignfig}.  A likely cause of this behavior could be the increase of the PSF size (i.e.~full width at half maximum) with wavelength \citep{wfc3handbook}, causing the PSF to be increasingly undersampled in $F110W$, but otherwise we found that neither the alignment rms nor the per-cluster difference between $F110W$ and $F160W$ alignment rms correlated with exposure time or total number of stars used for alignment, demonstrated by -0.1$<$$\rho$$<$0 and p-values $>$0.5 from a Spearman rank correlation.

\subsection{PSF Photometry \label{psfsect}}

Recently, a detailed description of the use of \texttt{Dolphot} to generate deep, multi-band, multi-instrument photometric catalogs was presented by \citet{phat} in the context of the Panchromatic Hubble Andromeda Treasury (PHAT).  In that study, extensive artificial star tests were performed to select an optimal set of \texttt{Dolphot} PSF photometry parameters and subsequent cuts on photometric quality parameters (output by \texttt{Dolphot}) to optimize their CMDs given the crowded nature of many of their fields.  However, rather than adopt their best values wholesale, we independently validate their choice of photometry parameter values  because the nature of our targets and observation strategy differ from theirs in several critical ways.  
First, although some of the PHAT fields are extremely crowded, they are much less severely plagued by saturated stars and their diffraction spikes.  This is due to both the more distant nature of their targets, as well as the moderate Galactic latitude of their target sightlines through the Galaxy.  Conversely, all of our target GGCs are more nearby and located close to the Galactic plane, increasing the incidence of both saturated cluster giants and saturated foreground disk and/or bulge stars.  Second, our observation strategy differs from that of PHAT since the optical and near-infrared observations of our target clusters attain increasingly different photometric depths for more extincted clusters.  Furthermore, our observations are obtained over a single orbit per instrument so that large dithers and
a variety of roll angles are not available, for instance, to rederive distortion solutions.

Given these differences, we chose to validate the parameters used for PHAT by performing a series of \texttt{Dolphot} photometry runs and artificial star tests 
where parameters were varied, both in turn and in combination, between the values recommended in the \texttt{Dolphot} manual and those used for PHAT.  
Based on these tests, the final set of \texttt{Dolphot} parameters which we adopted is a mix of the default values and the PHAT values, and all parameters for which we employed values differing from
their defaults are listed in Table \ref{doltab}.

\begin{deluxetable}{lcc}
\tablecaption{Modified \texttt{Dolphot} Parameters \label{doltab}}
\tablehead{
\colhead{Parameter} & \colhead{WFC3/IR} & \colhead{ACS/WFC}
}
\startdata
\texttt{img\_RAper} & 3 & 4 \\
\texttt{img\_RChi} & 1.5 & 2.0 \\
\texttt{img\_RSky} & 8 20 & 15 35 \\
\texttt{Force1} & 1 & 1 \\
\texttt{FitSky} & 2 & 2 \\ 
\texttt{Align} & 4 & 4 \\ 
\texttt{RCombine} & 1.415 & 1.415 \\ 
\texttt{PosStep} & 0.1 & 0.1 \\ 
\texttt{SigFind} & 3 & 3 \\ 
\texttt{dPosMax} & 2.5 & 2.5 \\ 
\enddata
\end{deluxetable}

\subsection{Photometric Quality Cuts \label{qualcutsect}}

Our primary science goals, including the measurement of cluster luminosity and mass functions,
necessitate that the raw catalogs output by \texttt{Dolphot} are effectively cleaned of spurious detections.  The \texttt{Dolphot} catalogs give several diagnostic parameters for each detection, including sharpness \texttt{sharp}, roundness \texttt{round}, the crowding \texttt{crowd} (the brightness increase in magnitudes which would result from contamination by neighboring stars), and a goodness-of-fit parameter \texttt{chi}, which the \texttt{Dolphot} manual warns may not be trustworthy in an absolute sense (although we employ it in a relative sense below).  For each detected source, each of these parameters is given as a global value per source from the combined photometry across all images, and individual per-filter values are also output for each filter in which a source is detected.  We employ two different sets of photometric quality criteria depending on instrument, one for the ACS/WFC $F606W$ photometry and one for the WFC3/IR photometry, and apply these cuts in each filter separately using the per-filter diagnostic parameters.  This approach proved superior to the use of the global diagnostic parameters reported by \texttt{Dolphot} because of the differences in pixel scale between instruments as well as varying difference in photometric depth between instruments as a function of crowding and line of sight extinction among the target clusters.

For the ACS/WFC imaging, we found that fixed values were sufficient to exclude spurious detections, including diffraction spikes from saturated stars, while exploiting its improved spatial resolution over WFC3/IR to detect real sources in crowded cluster cores.  These fixed values varied slightly from cluster to cluster, based on inspection of the spatial and CMD loci of kept and rejected sources, particularly ensuring that the maximum allowed value of \texttt{crowd} is set sufficiently low to exclude diffraction spikes and cosmetic defects while retaining as many true stellar detections as possible.  

\begin{figure}
\plotone{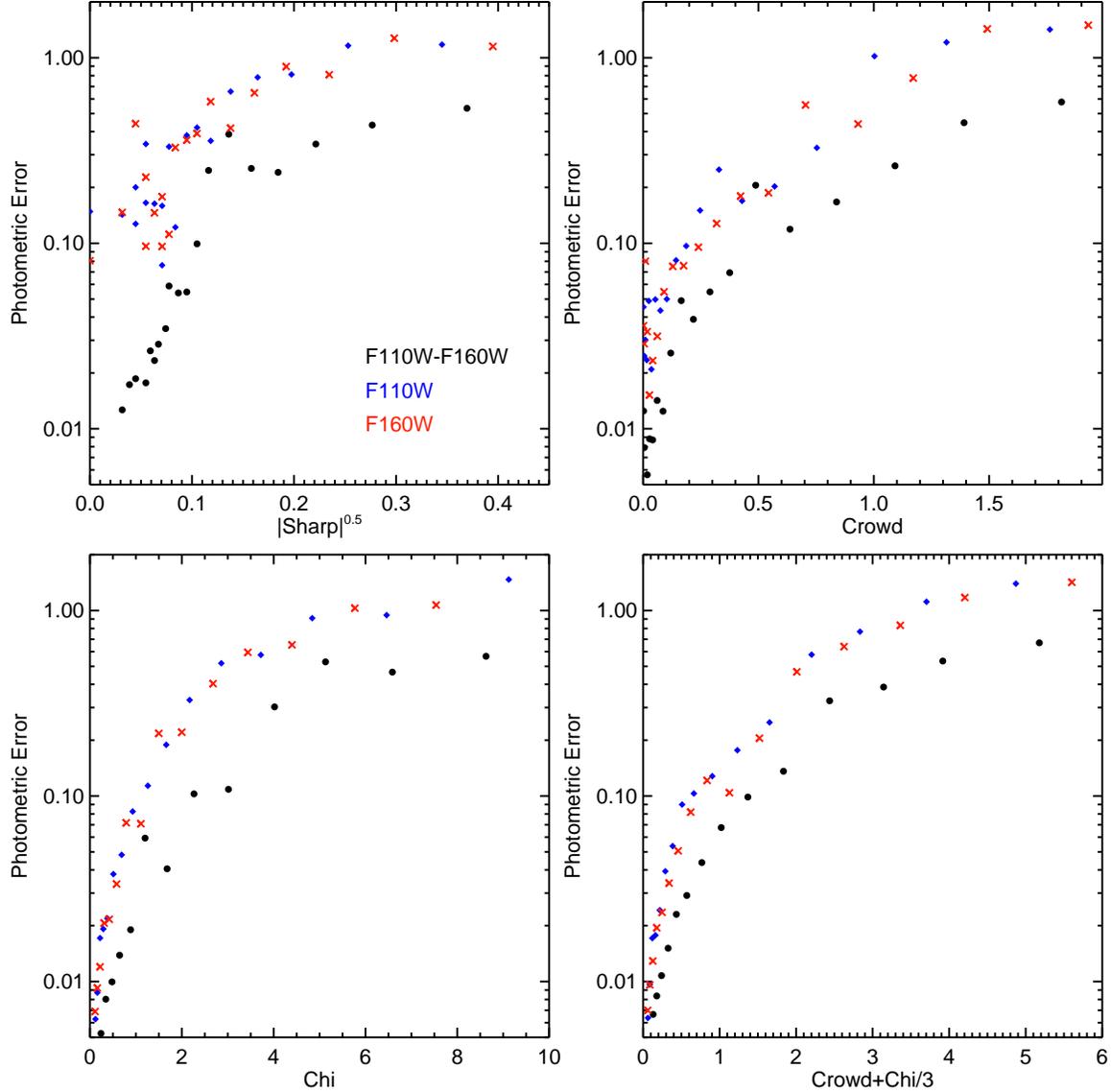}
\caption{Photometric errors from artificial star tests as a function of several diagnostic parameters, to demonstrate their effectiveness in isolating the most well-measured detections.  Median errors in $(F110W-F160W)$ color is illustrated using black filled points, and $F110W$ and $F160W$ magnitude is shown using blue diamonds and red crosses respectively.  Comparison of the four panels illustrates that \texttt{crowd}+(\texttt{chi}/3) maps most clearly to photometric error and is therefore the most useful parameter to selectively eliminate poorly-measured sources with large errors.}   
\label{paramVerrfig} 
\end{figure}

For WFC3/IR detections, we found that combinations of fixed cuts in the diagnostic parameters or sigma clipping with magnitude did not yield the most efficient rejection of artifacts.  Rather, a combination of the \texttt{crowd} and \texttt{chi} parameters yielded the most straightforward constraints on photometric quality, such that a maximum value of \texttt{crowd}+(\texttt{chi}/3) ranging from 1.3 to 2.0 (again, based on inspection of spatial and CMD distributions as well as visual inspection of the images) was applied to each cluster.  The ability of this parameter to most directly select well-measured sources is illustrated in Fig.~\ref{paramVerrfig}.  There, we compare the ability of \texttt{crowd}+(\texttt{chi}/3) to \texttt{crowd}, \texttt{chi} and $\vert$\texttt{sharp}$\vert$ to most directly select the stars with the smallest photometric errors.  Artificial stars (see Sect.~\ref{artstarsect}) were sorted into equally populated bins in each parameter on the horizontal axis, and the median photometric error given by the artificial star tests was plotted for each bin on the vertical axis.  Median errors are shown in terms of each of the two WFC3/IR magnitudes as well as $(F110W-F160W)$ color, and it is clear that of the four diagnostic parameters, \texttt{crowd}+(\texttt{chi}/3) shows the most clear monotonic trend with photometric error or, in other words, the best discriminating power to select well-measured stars.
Our photometric quality cuts for each filter are summarized in Table \ref{qualcuttab}.

\begin{deluxetable}{c}
\tablecaption{Photometric Quality Cuts \label{qualcuttab}}
\tablehead{
\colhead{All Filters} 
}
\startdata
Object Type = 1 \\
Quality Flag $\leq$2 \\
\hline
$F606W$ \\
\hline
$\vert$\texttt{sharp}$\vert$$\leq$0.3 \\
0.3$\leq$max(\texttt{crowd})$\leq$1.0 \\
2$\leq$min(S/N)$\leq$10 \\
\hline
$F110W,F160W$ \\
\hline
1.3$\leq$max(\texttt{crowd} + \texttt{chi}/3)$\leq$2 \\
5$\leq$min(S/N)$\leq$10 \\
\enddata
\end{deluxetable}

\begin{figure}
\plotone{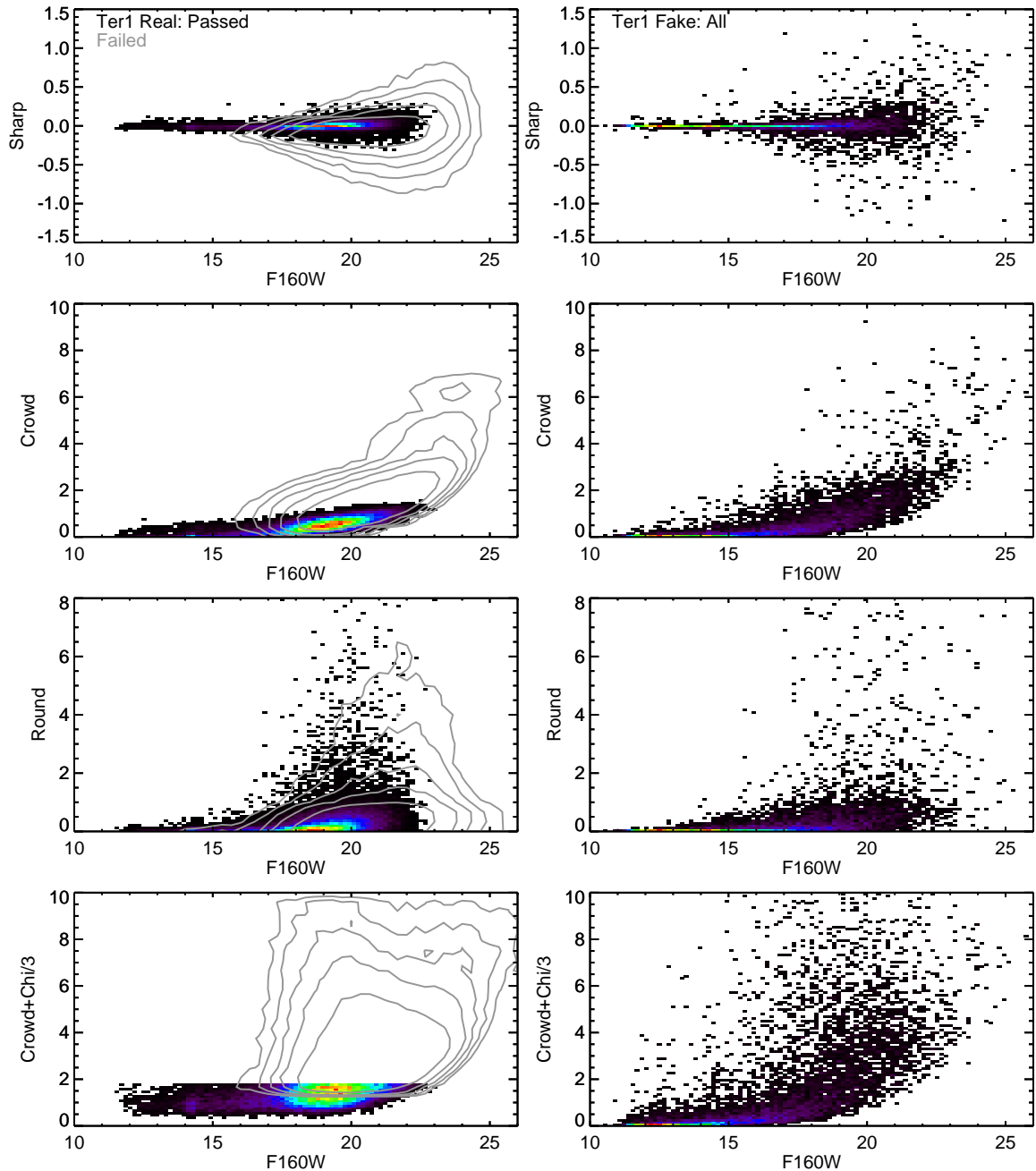}
\caption{The diagnostic parameters \texttt{sharp}, \texttt{crowd}, \texttt{round} and \texttt{crowd}+\texttt{chi}/3 as a function of F160W magnitude.  \textbf{Left Column}: Density plots of stars in our observed catalog for the target cluster Terzan 1 which passed our quality cuts (colored on a linear density scale).  The grey contours indicate the loci of stars in the observed catalog which failed our quality cuts, and are plotted at intervals of (0.0125,0.025,0.05,0.1,0.25) times the maximum value to illustrate the full region where rejected sources lie with respect to each parameter.  \textbf{Right Column:} Density plots of all artificial stars, shown on the same scale as the left column.  Areas which are empty in these plots should be occupied only by spurious detections in the raw observed catalog.  The difference in the density distribution between the two columns is partially by construction, since artificial stars were given input $F160W$ magnitudes with a flat distribution (see Sect.~\ref{artstarsect}).}   
\label{diagparamfig} 
\end{figure}

Additional verification of photometric quality cuts can be gained from the artificial star tests themselves.  While they quantify photometric incompleteness due to application of the cuts, the distribution of the artificial stars with respect to the various diagnostic parameters also indicates where true stellar detections are expected \textit{not} to lie.  Values of the diagnostic parameters seen only in the observed catalogs but not output by the artificial star tests are therefore indicators of where only false detections exist.  The ability of our cuts to retain stars falling in the (diagnostic) parameter space occupied primarily by artificial stars while rejecting sources falling outside this space is illustrated for the example case of 
Terzan 1\footnote{Throughout this section, we use Terzan 1 as an example case because star counts over our target cluster parallel fields in the magnitude range where they are all $>$90\% complete
(14$\leq$$F160$$\leq$16)
indicate that Terzan 1 has the highest field star density, and therefore represents a relatively pathological case for the effects of crowding.  At the other extreme, NGC 6638 has the lowest parallel field stellar density in this magnitude range (more than 7 times lower than Terzan 1).} in Fig. \ref{diagparamfig}.  There, the left column shows density plots of various diagnostic parameters versus $F160W$ magnitude for real stars which passed our cuts, and the right column shows density plots of the same parameters, but for all artificial stars (not just those passing the cuts).  The loci of observed sources which failed our cuts, on the other hand, are indicated by the grey contours in the left column.  While it is apparent that some stellar sources will be rejected, the similarity in the two distributions demonstrates that the criterion of \texttt{crowd}+\texttt{chi}/3 does a reasonable, and intentionally somewhat conservative, job of selecting well-measured sources.  While the density distributions between the observed and artificial star catalogs appear somewhat dissimilar in high density regions, this is by construction, and is simply a result of the flat input magnitude distribution assigned to the artificial stars.  Moreover, while our use of a fixed cut in \texttt{crowd} + \texttt{chi}/3 effectively imposes a faint magnitude limit, the artificial star tests reveal that at such faint magnitudes, completeness is low, photometric errors are large, and there is a substantial bias \textit{in the mean} between input and recovered magnitudes (see Sects.~\ref{artstarsect} and \ref{compreducsect}).

The fraction of initially detected sources passing our photometric quality cuts is generally fairly low, particularly in the WFC3/IR filters, where between 8.7\% and 21.8\% of the sources in the WFC3/IR field of view passed our quality cuts in both $F110W$ and $F160W$.  However, such values are not uncommon in similar applications of \texttt{Dolphot} to produce WFC3/IR photometry in crowded fields via the use of \texttt{Force1}=1 while requiring a thorough rejection of spurious detections \citep[e.g.][]{m4bdir}.  Furthermore, the possibility that our cuts are overly aggressive is unlikely based on the high ($>$90\%) completeness of the WFC3/IR photometry over a range of at least $\sim$4.5 mag for all target clusters (see Sect.~\ref{artstarsect}).  Meanwhile, the fraction of sources passing our cuts in both WFC3/IR filters but failing in $F606W$ showed large variations (ranging from 7\% to 51\%) which correlate, predictably, with the amount of total and differential extinction seen in the cluster CMDs.  Of the sources detected in $F606W$, less variation is seen, and the fraction passing our cuts ranges from 15.4\% to 38.5\%.

\subsection{Artificial Star Tests and Completeness \label{artstarsect}}

Artificial star tests are used to quantify incompleteness as well as photometric and astrometric uncertainties as a function of position, color and magnitude.  Artificial stars are assigned input magnitudes drawn from a flat luminosity function in $F160W$, and for each of these stars, its color in $(F606W-F160W)$ and $(F110W-F160W)$ is randomly drawn from the colors of real stars within $\pm$1 mag\footnote{At the extreme bright and faint ends of the observed magnitude range, this value is increased up to as much as 2 mag in order to sample the full range of plausible input colors and magnitudes spanning our detection limits.} in $F160W$.  The positions of the artificial stars are concentrated towards the cluster center as are the real stars, so that the completeness ascertained from the full ensemble of artificial stars over any magnitude/color range is a reasonable proxy for the true (radially integrated) value, although we do not assume this to be true in a quantitative sense.  In practice, the artificial star positions are drawn randomly from a subset of relatively bright ($F606W$$\lesssim$21) real stars before applying small random offsets in position, magnitude and color to ensure adequate input spatial and color-magnitude coverage over the entire observed distribution.  Artificial stars are inserted one at a time and PSF photometry is performed identically as for real stars, so that each artificial star is affected by any real neighbors present in the science images, but each artificial star is not affected by other artificial stars.  The resulting 50\% and 90\% bright and faint completeness limits after 
application of the photometric quality cuts described in Sect.~\ref{qualcutsect} are listed for
each filter, for both the primary and parallel fields, in Table \ref{completetab}.

\begin{splitdeluxetable*}{lccccccccccccBcccccccc}
\tablecaption{Primary and Parallel Field Completeness Limits \label{completetab}}
\tabletypesize{\footnotesize} 
\tablehead{
 \colhead{} & \multicolumn{4}{c}{F606W Primary} & \multicolumn{4}{c}{F110W Primary} & \multicolumn{4}{c}{F160W Primary} &  \multicolumn{4}{c}{F110W Parallel} & \multicolumn{4}{c}{F160W Parallel} \\
\colhead{} & \colhead{$C^{90}_{min}$} & \colhead{$C^{90}_{max}$} & \colhead{$C^{50}_{min}$} & \colhead{$C^{50}_{max}$} & \colhead{$C^{90}_{min}$} & \colhead{$C^{90}_{max}$} & \colhead{$C^{50}_{min}$} & \colhead{$C^{50}_{max}$} & \colhead{$C^{90}_{min}$} & \colhead{$C^{90}_{max}$} & \colhead{$C^{50}_{min}$} & \colhead{$C^{50}_{max}$} & \colhead{$C^{90}_{min}$} & \colhead{$C^{90}_{max}$} & \colhead{$C^{50}_{min}$} & \colhead{$C^{50}_{max}$} & \colhead{$C^{90}_{min}$} & \colhead{$C^{90}_{max}$} & \colhead{$C^{50}_{min}$} & \colhead{$C^{50}_{max}$} 
}
\startdata
BH261 & 15.607 & 23.454 & 13.975 & 26.704 & 11.902 & 18.550 & 11.384 & 21.654 & 13.117 & 18.962 & 12.699 & 22.293 & 11.654 & 15.773 & 11.243 & 21.773 & 13.542 & 18.916 & 12.863 & 22.632 \\
Djorg2 & 15.446 & 22.713 & 14.802 & 26.295 & 11.700 & 17.118 & 11.260 & 20.048 & 13.151 & 18.408 & 12.640 & 21.082 & 11.934 & 16.831 & 11.219 & 19.700 & 13.705 & 16.928 & 12.971 & 20.549 \\
FSR1735 & 16.251 & 26.517 & 15.487 & 27.673 & 11.816 & 18.912 & 11.350 & 21.427 & 13.183 & 20.170 & 12.745 & 23.027 & 11.934 & 17.981 & 11.313 & 23.133 & 13.770 & 19.100 & 13.014 & 24.397 \\
HP1 & 16.122 & 23.267 & 15.487 & 26.243 & 11.829 & 17.854 & 11.276 & 19.658 & 13.043 & 18.912 & 12.611 & 20.725 & 12.144 & 17.161 & 11.525 & 20.027 & 13.440 & 18.245 & 12.907 & 21.186 \\
NGC6540 & 15.396 & 22.596 & 14.074 & 26.398 & 11.689 & 17.391 & 11.116 & 20.198 & 13.132 & 18.147 & 12.684 & 21.066 & 12.028 & 16.992 & 11.192 & 20.280 & 13.528 & 17.460 & 12.872 & 21.218 \\
NGC6569 & 15.197 & 21.617 & 14.597 & 25.438 & 11.864 & 17.135 & 11.426 & 19.883 & 12.953 & 18.079 & 12.634 & 20.679 & 12.052 & 17.733 & 11.424 & 23.647 & 13.670 & 18.361 & 12.953 & 24.532 \\
NGC6638 & 14.318 & 20.703 & 13.833 & 24.871 & 11.715 & 16.141 & 11.187 & 19.030 & 13.066 & 16.929 & 12.674 & 19.925 & 11.773 & 18.427 & 11.197 & 23.340 & 13.851 & 19.244 & 13.169 & 24.731 \\
Pal6 & 15.972 & 23.658 & 15.288 & 26.516 & 11.721 & 16.194 & 11.103 & 19.178 & 13.283 & 17.711 & 12.692 & 20.559 & 12.071 & 14.734 & 11.403 & 19.827 & 13.537 & 17.066 & 12.861 & 20.896 \\
Ter1 & 16.079 & 23.907 & 15.155 & 26.740 & 11.736 & 15.599 & 11.205 & 18.385 & 13.266 & 17.245 & 12.649 & 19.905 & 11.889 & 16.776 & 11.312 & 19.717 & 13.605 & 16.969 & 12.719 & 21.317 \\
Ter2 & 16.396 & 24.135 & 15.919 & 27.178 & 11.726 & 17.444 & 11.151 & 19.859 & 13.116 & 18.516 & 12.644 & 21.361 & 11.907 & 16.010 & 11.268 & 20.315 & 13.570 & 18.275 & 12.824 & 21.515 \\
Ter4 & 16.501 & 25.614 & 15.801 & 26.853 & 11.780 & 17.668 & 11.189 & 20.057 & 13.399 & 18.889 & 12.716 & 21.595 & 11.933 & 15.829 & 11.263 & 20.604 & 13.462 & 18.003 & 12.627 & 22.052 \\
Ter6 & 16.798 & 24.763 & 15.831 & 26.877 & 11.758 & 16.253 & 11.200 & 19.429 & 13.243 & 17.797 & 12.479 & 20.881 & 11.864 & 16.157 & 11.300 & 19.964 & 13.542 & 18.025 & 12.905 & 21.567 \\
Ter9 & 16.767 & 23.866 & 15.579 & 27.104 & 11.563 & 16.808 & 11.102 & 18.916 & 13.127 & 18.447 & 12.585 & 20.984 & 11.922 & 16.989 & 11.350 & 20.283 & 13.492 & 17.587 & 12.911 & 21.545 \\
Ter10 & 16.601 & 24.996 & 16.000 & 27.377 & 11.739 & 17.767 & 11.212 & 20.276 & 13.345 & 18.865 & 12.658 & 21.200 & 11.889 & 16.518 & 11.357 & 20.603 & 13.591 & 16.768 & 12.826 & 21.880 \\
Ter12 & 16.519 & 25.086 & 15.763 & 27.452 & 11.677 & 17.181 & 11.202 & 19.829 & 13.255 & 18.668 & 12.775 & 21.480 & 12.107 & 16.631 & 11.443 & 21.090 & 13.632 & 19.012 & 12.971 & 22.597 \\
Ton2 & 16.219 & 23.665 & 15.504 & 27.152 & 11.783 & 17.834 & 11.314 & 20.500 & 13.227 & 18.884 & 12.696 & 21.971 & 12.149 & 18.678 & 11.401 & 22.860 & 13.649 & 18.527 & 13.000 & 24.683 \\
\enddata
\tablecomments{For each filter, $C^{90}_{min}$ and $C^{90}_{max}$ denote the 90\% bright and faint completeness limits, while $C^{50}_{min}$ and $C^{50}_{max}$ denote the 50\% bright and faint completeness limits.  All values are magnitudes in the Vegamag photometric system native to each instrument as described in Sect.~\ref{cmdsect}.}
\end{splitdeluxetable*}

\subsection{Comparison of Reduction Strategies \label{compreducsect}}

\begin{figure}
\plottwo{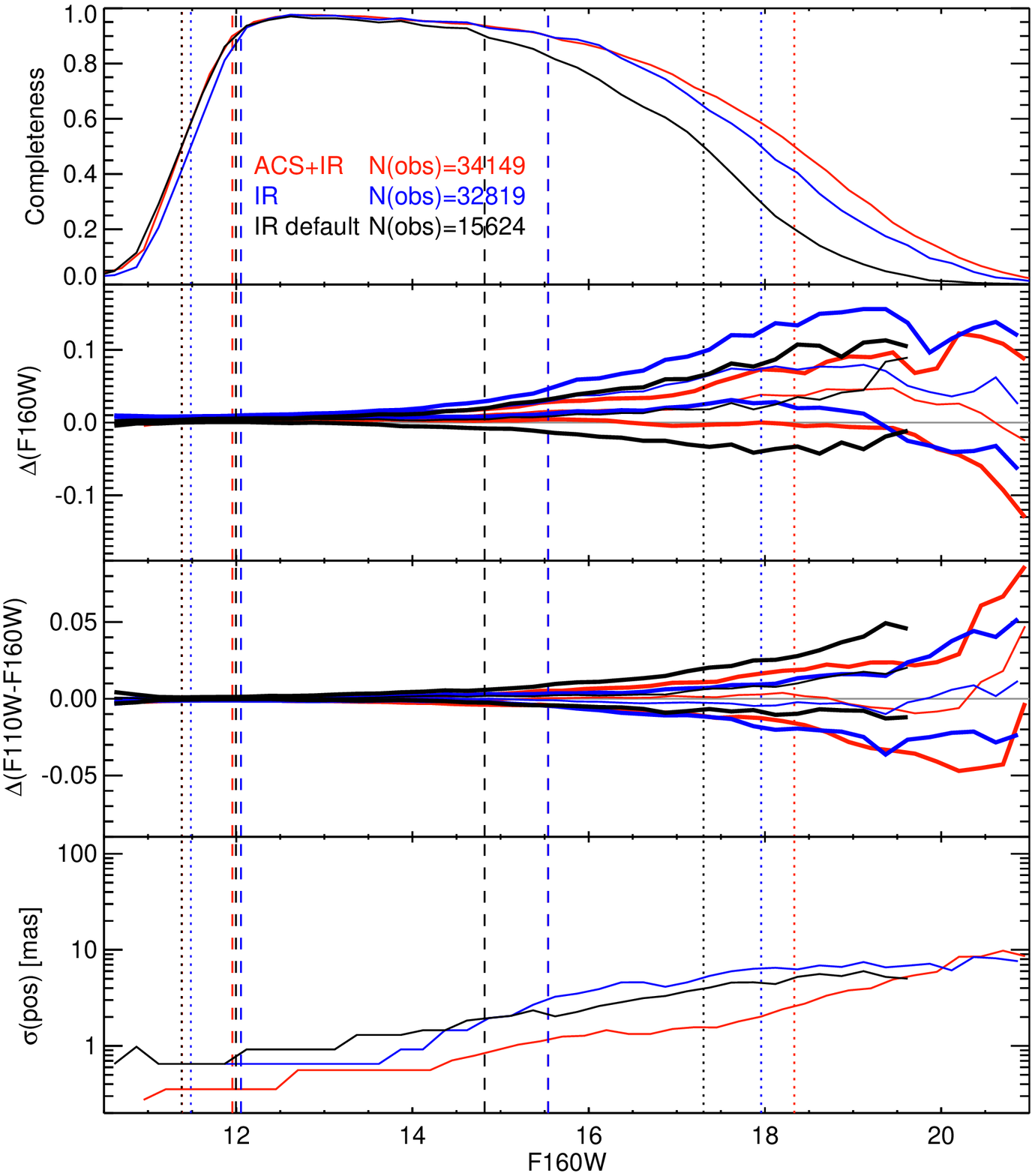}{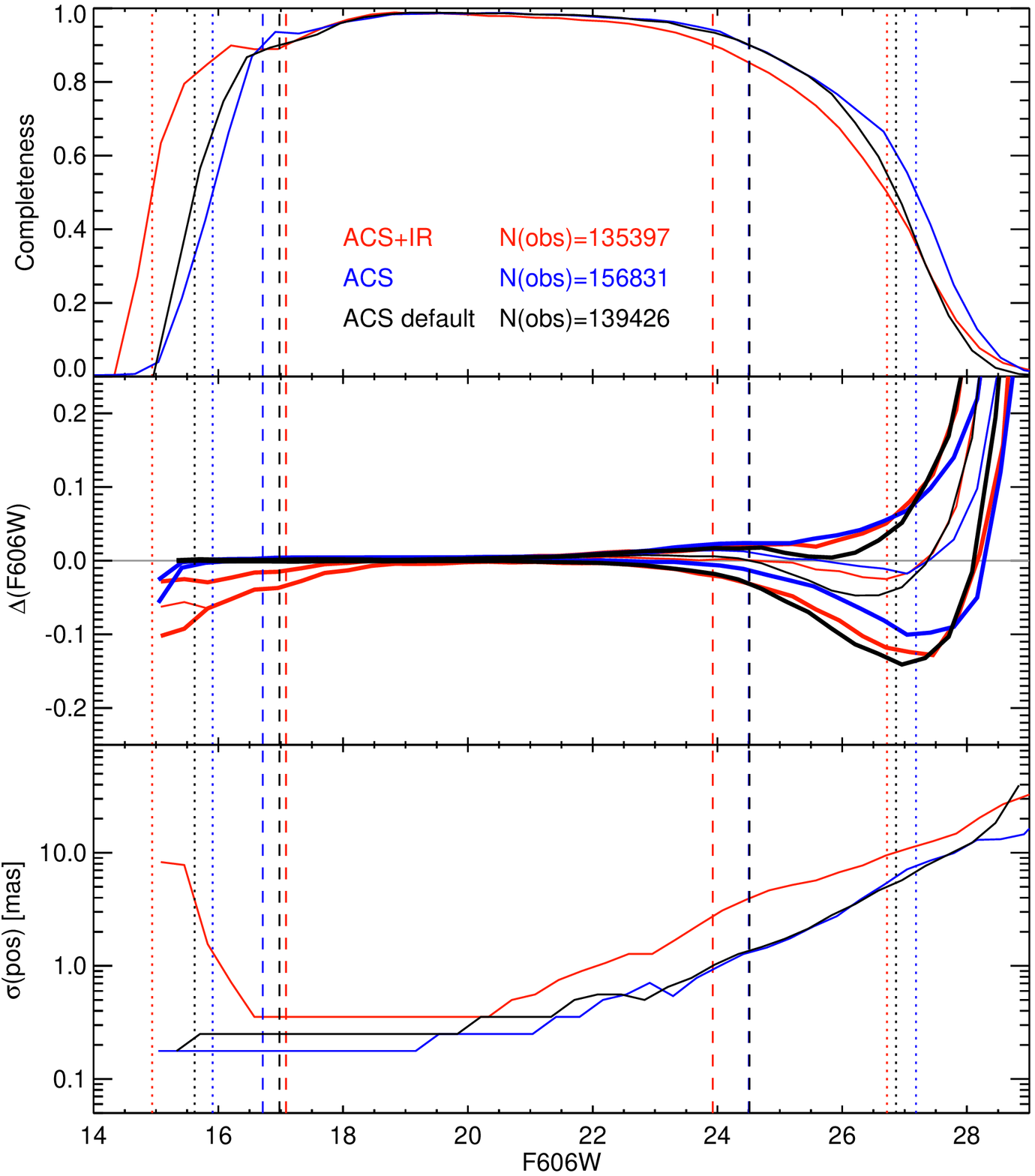}
\caption{Comparison of photometric and astrometric accuracy from different \texttt{Dolphot} data reduction strategies using artificial star tests, for WFC3/IR images (left column) and ACS/WFC images (right column) of Terzan 1.  In each plot, we compare three reduction strategies: The default recommended \texttt{Dolphot} preprocessing and photometry parameters (black), our optimized parameters for each instrument separately (blue), and optimized parameters used to perform simultaneous photometry of ACS and IR images together (red).  For each of the three strategies, 50\% and 90\% completeness limits are shown as dotted and dashed vertical lines respectively. \label{comprunsfig}}    
\end{figure}

We have also used artificial star tests to quantify the advantages and disadvantages of different reduction strategies.  The completeness limits (Table \ref{completetab}) and
CMDs we present (Figs.~\ref{cmdfig1}-\ref{cmdfiglast}) employ simultaneous photometry of optical and IR images
as described in Sect.~\ref{preprocsect}-\ref{psfsect}.  However, we have also performed IR-only and 
ACS-only full reductions in order to compare and assess, for example, whether a particular strategy improves or degrades astrometric versus photometric accuracy.  Specifically, for the set of IR images, we performed two \texttt{Dolphot} runs independently of the simultaneous optical+IR run already described.  The first was performed with the default preprocessing and parameter values recommended
in the \texttt{Dolphot} manual.  The second was performed with the optimized parameters
given in Table \ref{doltab}, but without the inclusion of the ACS/WFC images.  
In the left panel of Fig.~\ref{comprunsfig}, we plot several quantities as a function of
input $F160W$ magnitude for each of the three (including the original ACS+IR run) \texttt{Dolphot} runs.  In the top panel, we plot the completeness fraction, and 
also give the number of stars passing the photometric quality cuts described in Sect.~\ref{qualcutsect}.  In the second and third panels, we plot the median (thin line)
and $\pm$1$\sigma$ (thick lines) difference in $F160W$ magnitude and $(F110W-F160W)$ color.
In the bottom panel, we plot the positional error as a function of magnitude.  In a practical sense, the positional error given here is more appropriately viewed as a lower limit since \texttt{Dolphot} assumes the same distortion solution to place artificial stars on individual images as well as transform detections in these images to the
reference image frame.

As a guide, we have used the values in the top panel to indicate the 50\% and 90\% completeness limits for each of the three runs as dotted and dashed vertical lines.  In the right-hand panel of Fig.~\ref{comprunsfig}, we show the results of an analogous experiment for the ACS/WFC imaging, where we compare the ACS+IR reduction strategy to ACS-only \texttt{Dolphot} runs using both the default parameters and our optimized parameters.  Taking both panels of Fig.~\ref{comprunsfig} together, several results are apparent:

\begin{enumerate}

\item For any given run, the magnitude offset (bias) never exceeds zero at a
significant ($>$1$\sigma$) level, over all magnitudes within the 50\% completeness limits.  The only exception to this rule is the brightest stars in the ACS/WFC images, which are only recovered when photometered simultaneouly with WFC3/IR.  However, in this case, they show a median offset of $\sim$0.05 mag in $F606W$ and no offset in WFC3/IR magnitude or color.
 
\item Unsurprisingly, the positional accuracy of WFC3/IR detections is improved when photometered simultaneously with ACS/WFC.  Similarly, the positional accuracy of ACS/WFC is degraded with the inclusion of WFC3/IR.  However, even in this case, the internal positional errors for bright, unsaturated sources are about 0.5 mas, or 0.01 ACS/WFC pix, roughly consistent with what has been
feasible by \citet{hstpromo1} per epoch with a similar ACS/WFC observing strategy.

\item In the default run for WFC3/IR, and to a lesser extent, the IR-only run, a smaller number of sources are detected and completeness is lower at a given $F160W$ magnitude.  At face value, this could simply be suggesting that our quality cuts are overly harsh in these cases.  However, for these runs, the photometric errors in color and magnitude are not smaller, and are in fact larger, than those from the ACS+IR photometry at a given magnitude, indicating that the ACS+IR photometry does in fact recover a greater number of sources with smaller photometric errors.

\item The ACS-only runs appear to provide photometry which is of equal quality and greater astrometric precision than the ACS+IR runs, with the exception of sources near the saturation limit mentioned above.  However, significant magnitude offsets are seen in all cases faintward of the 50\% completeness limit which we were unable to further mitigate using photometric quality cuts.  Therefore, the $F606W$ photometry should not be considered trustworthy in a quantitative sense at such faint magnitudes, even in light of photometric errors of $>$0.1 mag.

\end{enumerate}

The results we present in Sect.~\ref{cmdsect} are based on a simultaneous \texttt{Dolphot} reduction of the ACS and IR
images together, since the artificial star tests using this strategy (shown in red in Fig.~\ref{comprunsfig})
reveal the best combination of dynamic range and photometric precision compared to the other strategies for
the purpose of constructing clean, multi-band photometric catalogs.  However, it is also clear that specific
science goals may benefit
(for example, in positional accuracy) from tailored alternate data reduction strategies.

\subsection{Astrometric Calibration}

The final cleaned, distortion corrected photometric catalogs are astrometrically calibrated using our 2MASS-calibrated VVV PSF photometry catalogs.  Given the large and variable optical extinction towards many of our target clusters, near-IR photometry is needed to provide a sufficient number of calibrating stars in the field of view of our observations.  Furthermore, our VVV PSF catalogs provide significant advantages over 2MASS in terms of both photometric depth and spatial resolution \citep{maurohb,cohen6544,cohenvvv}, rendering them vastly more compatible with the \textit{HST} imaging obtained here.  Briefly, the astrometric calibration for each target cluster was performed as follows: First, only bright ($K_{S}$$\lesssim$15), unsaturated sources detected in multiple bands in the VVV catalogs are considered for use as astrometric standards.  Given offsets of over 1 arcsec seen in the header WCS of the \textit{HST} reference image, we use a simple matching algorithm with a 5$\sigma$ clip (as described in Sect.~\ref{preprocsect} but with a more generous convergence criterion of 0.01$\arcsec$) to cross reference distortion-free pixel coordinates of \textit{HST} detections with (2MASS-calibrated) RA and Dec of VVV detections.  Lastly, the IRAF \texttt{ccmap} and {cctran} tasks are used (rejecting outliers with a 3$\sigma$ clip) to calculate and apply the final astrometric solution.  The number of stars used as astrometric calibrators varied between 605 and 1995 among our target clusters, and the astrometric rms deviation among these calibrators was $\leq$0.025$\arcsec$ per coordinate in all cases. 

\section{Color-Magnitude Diagrams \label{cmdsect}}

Here, we present the observed CMDs of all 16 target clusters, in several different
color-magnitude planes.  The magnitudes and colors reported here have had aperture corrections applied by \texttt{Dolphot} and are placed on the Vegamag system using ACS/WFC zeropoints and encircled energy corrections from \citet{bohlin} and WFC3/IR values posted on the STScI website\footnote{\url{www.stsci.edu/hst/wfc3/analysis/ir\_phot\_zpt}}.  For each CMD, we use dashed and solid grey
lines to illustrate the radially integrated 50\% and 90\% completeness limits.  For each of the four color-magnitude planes, the scales of the color and magnitude axes are kept fixed from cluster to cluster to facilitate a comparison (although the values of the axes may shift).
We now move on to a discussion of the individual clusters, summarizing previous literature
studies in each case.  This is done primarily to emphasize the lack of previous studies
of these clusters, but also to serve as a basis for comparison of cluster distances, 
reddenings and metallicities.  Remarkable effort has been made to place these values on a self-consistent footing to the extent possible by \citet{h96}, which is the only catalog to do so.  However, their reliance (by necessity) on uncorrected optical photometry (to obtain distances, structural parameters, and in some cases metallicities), along with the assumption of a constant, standard optical extinction law, may cause systematic biases in the resulting values.  For this reason, we revisit the earlier studies of each target cluster as completely as practically feasible, so that their methodology can be examined on a case-by-case basis.

\subsection{BH 261 = AL 3 = ESO 456-78}

\begin{figure}
\plotone{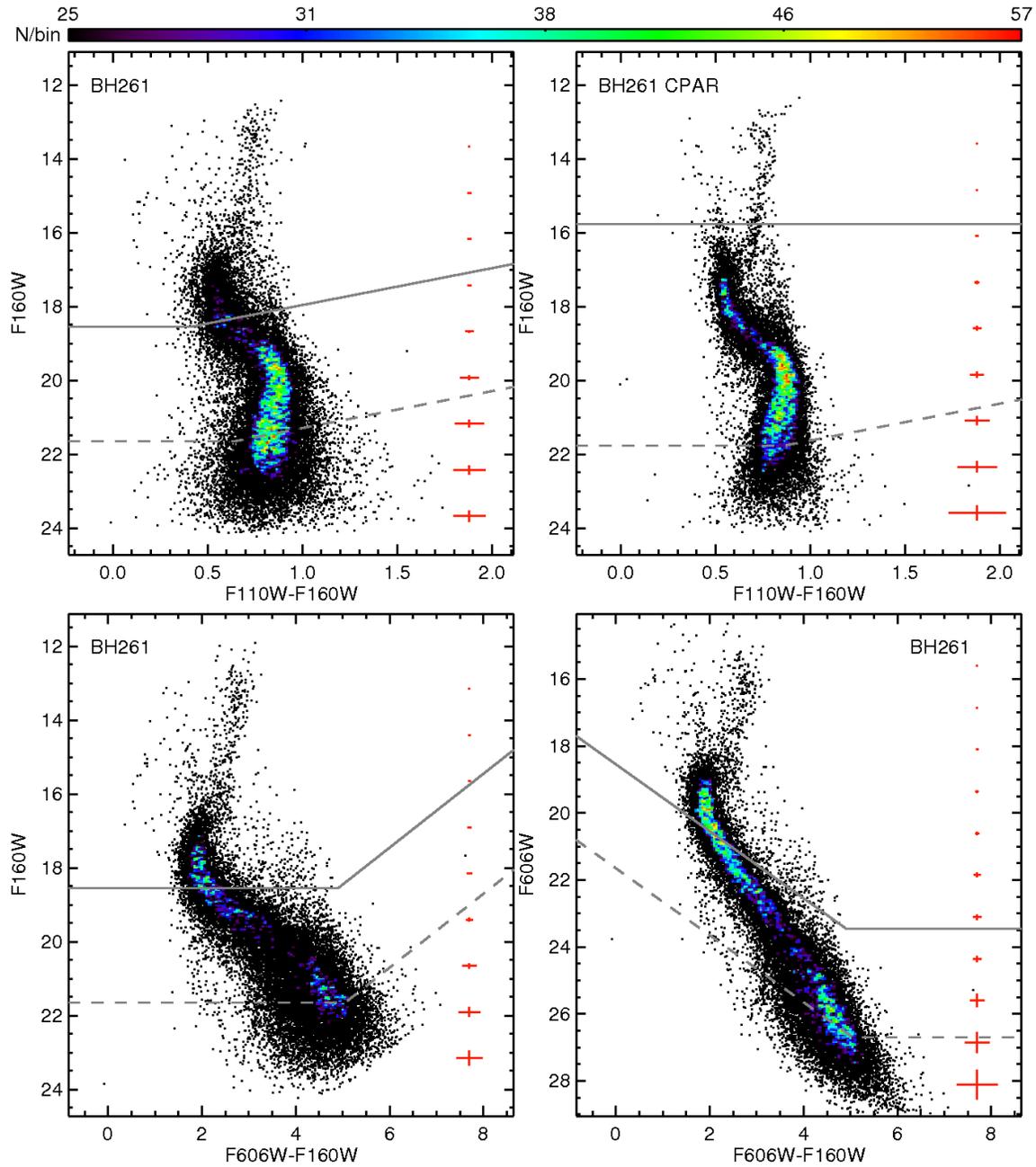}
\caption{CMDs of the target cluster BH261 in several color-magnitude planes, including the parallel
field in the upper right panel.  The cluster CMDs in the other three panels are constructed from 
simultaneous photometry of all primary field images, showing all stars passing our quality cuts
described in Sect.~\ref{qualcutsect}.  CMDs are color-coded according to the logarithmic color scale
shown in the colorbar at the top, and median photometric errors resulting from artificial star tests
are shown in red along the right hand side of each CMD.  The 50\% and 90\% radially integrated
faint completeness limits are indicated by dotted and solid grey lines respectively. The color and magnitude scales of both CMD axes in each panel are kept fixed in all subsequent figures to enable direct comparisons between clusters. \label{cmdfig1}}
\end{figure}

Essentially all that is known about this cluster photometrically is based on the discovery paper of \citet{bh261discov}, utilizing optical $BVI$ photometry, although a \citet{king62} profile fit to statistically cleaned 2MASS photometry was performed by \citet{sb2mass}.  Based on radial cuts to their photometric catalog, \citet{bh261discov} claim that the cluster is quite sparse with a blue HB, and obtain cluster parameters via a direct comparison of the RGB and HB sequences to those of the GGC M 5.  They assume a standard ($R_{V}$=3.1) extinction law to find $R_{\odot}$=6.5$\pm$0.5 kpc, $E(B-V)$=0.36$\pm$0.03 and a photometric metallicity estimate of $[\rm{Fe/H}]$=-1.3$\pm$0.25.  Proper motion cleaning by \citet{pmrossi} supports the blue HB morphology and they find a large proper motion in Right Ascension (4.77$\pm$0.46 mas/yr) compared to the background bulge/disk population, but as this cluster has no published radial velocities, they are unable to constrain its three-dimensional motion.  We confirm the sparse blue HB extending faintward to $F160W$$\sim$16 at $(F110W-F160W)$$\sim$0.1 and $(F606W-F160W)$$\sim$0.5 in Fig.~\ref{cmdfig1}.  As in the case of the majority of our target clusters, our imaging appears to extend well faintward of the lower main sequence knee (LMSK), seen near $F160W$$\sim$20 in this case.  However, careful measurement of field contamination is needed to reveal to what extent this feature is comprised of bulge field stars superimposed on the cluster main sequence, although in this case we may be seeing largely cluster members given that this cluster lies $>$5 degrees from the Galactic plane.

\subsection{Djorg 2 = ESO 456-SC38}

\begin{figure}
\plotone{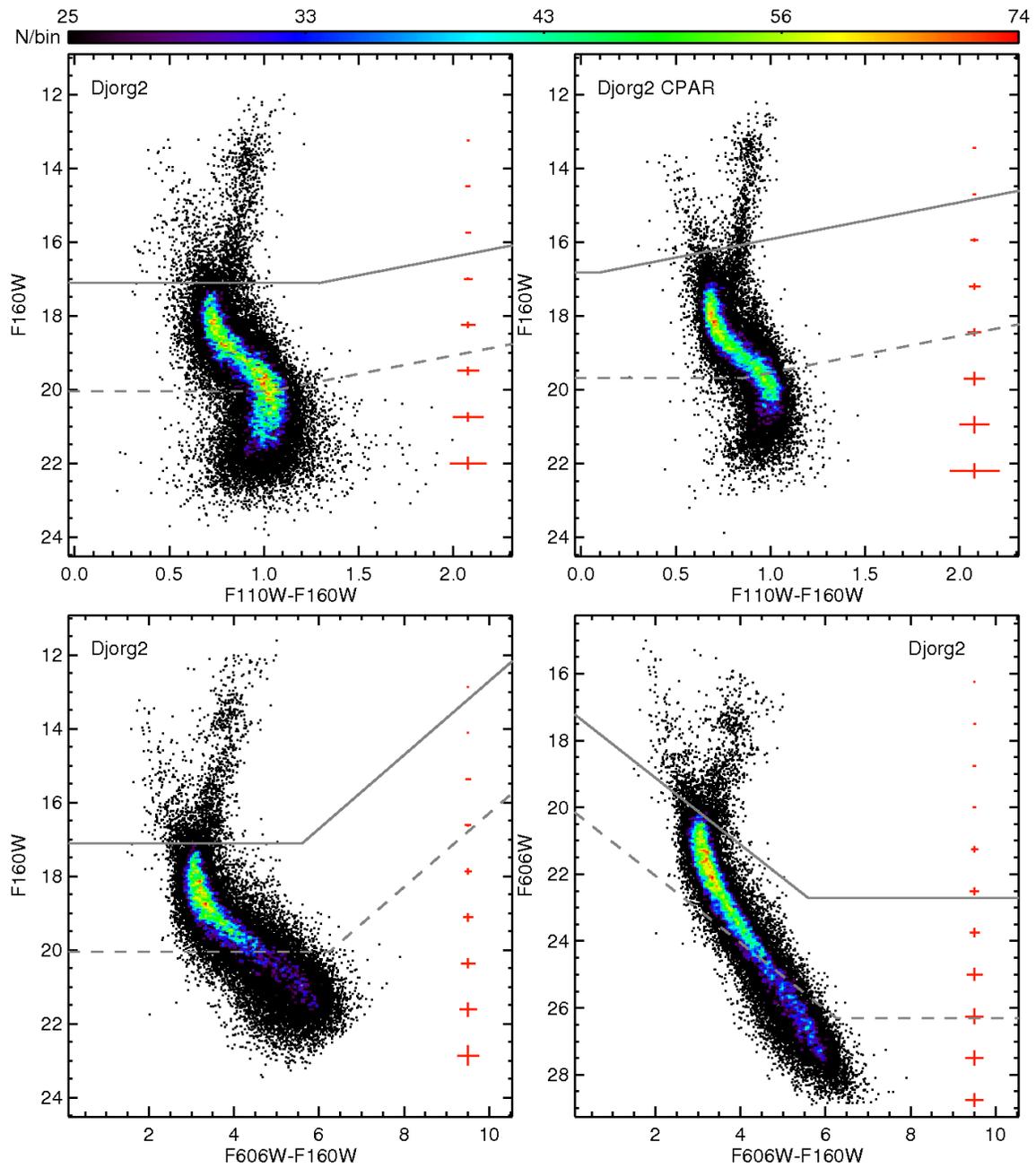}
\caption{As in Fig.~\ref{cmdfig1}, but for Djorg 2.
\label{cmdDjorg2}}
\end{figure}

Several studies of this cluster give a range of parameters.  Based on a comparison of optical photometry to the GGC 47 Tuc, \citet{djorg2discov} give $R_{\odot}$=5.5$\pm$0.8 kpc, $E(B-V)$=0.89$\pm$0.08 and a photometric metallicity estimate of $[\rm{Fe/H}]$$\sim$=-0.5.  Meanwhile, \citet{v10} use near-IR photometry to obtain $R_{\odot}$=7.0 kpc, $E(B-V)$=0.94 and a photometric metallicity of $[\rm{Fe/H}]$=-0.65.  Based on low-resolution optical spectra of four radial velocity-selected members, \citet{brunofors2} report $[\rm{Fe/H}]$=-0.79$\pm$0.09, although a discrepancy of $\sim$0.7 dex between the values they obtain from empirical versus synthetic spectral libraries, in addition to the small number of members available, suggests that an independent spectroscopic metallicity estimate would be valuable in this case, especially given the evidence of a blue HB extending to $F160W$$\sim$16.5 in Fig.~\ref{cmdDjorg2}, unusual at such a high metallicity.  A proper motion of (3.08,2.00)$\pm$(0.29,0.34) mas/yr in RA and Dec respectively is given by \citet{pmrossi}.

\subsection{FSR 1735 = 2MASS GC-03}

\begin{figure}
\plotone{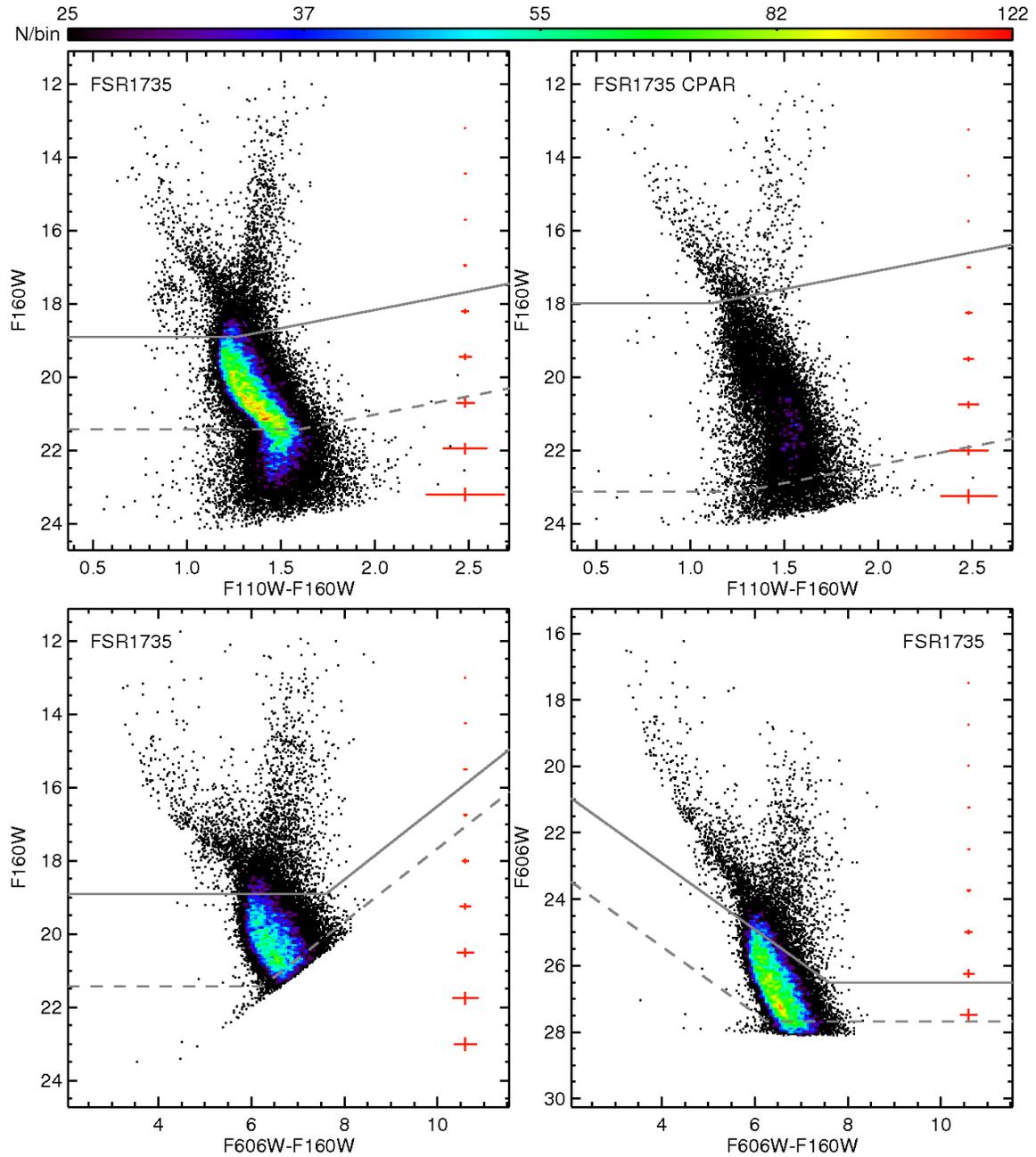}
\caption{As in Fig.~\ref{cmdfig1}, but for FSR 1735.
\label{cmdFSR1735}}
\end{figure}

The idea that this cluster candidate was in fact a GGC was put forth by \citet{fsr1735discov} based on near-IR photometry.  They give $R_{\odot}$=9.1$\pm$1.0 kpc based on the $K$-band magnitude of the red clump, using $A_{K}$=0.5$\pm$0.1, and a photometric metallicity of $[M/H]$=-0.8$\pm$0.1 based on the slope of the upper RGB.  Recently, \citet{fsr1735vvv} used subsequent near-IR photometry and spectroscopy to obtain $R_{\odot}$=10.8$\pm$0.9 kpc based on time series imaging of three ab-type RR Lyrae variables, and a spectroscopic metallicity of $[\rm{Fe/H}]$=-0.9$\pm$0.2 from five likely members.  Despite substantial differential reddening, evidence of a blue HB extending to $F160W$$>$18 is visible in Fig.~\ref{cmdFSR1735}, and comparison of the primary and parallel fields immediately reveals that in this case the primary source of contamination is the foreground Galactic disk rather than the bulge given the Galactic longitude of this cluster.

\subsection{HP 1 = ESO 455-11 = Gcl 67}

\begin{figure}
\plotone{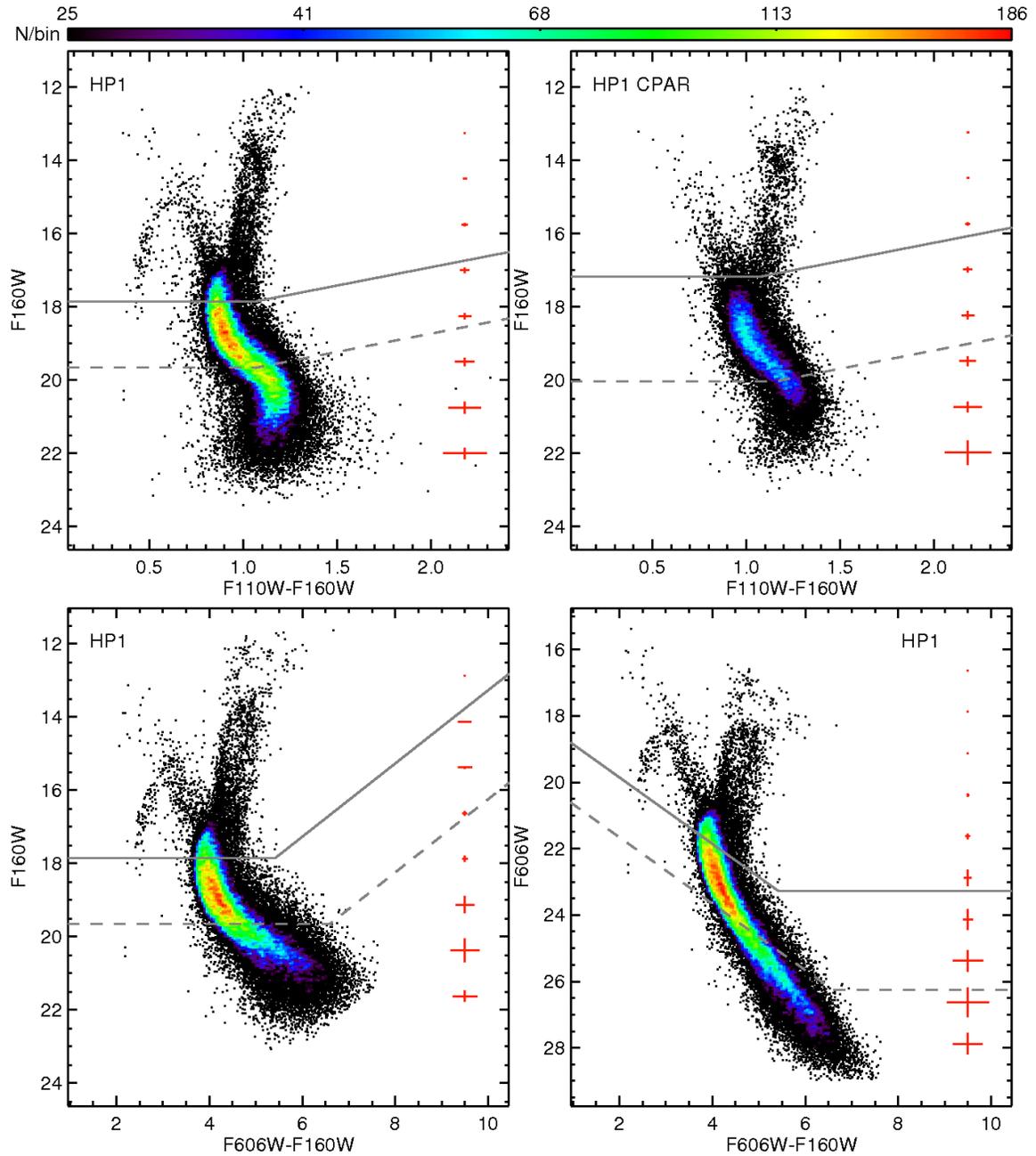}
\caption{As in Fig.~\ref{cmdfig1}, but for HP 1.
\label{cmdHP1}}
\end{figure}

Infrared photometry of this cluster and spectroscopy of six red giants was obtained by \citet{minniti95a}, 
finding $E(J-K)$=0.94$\pm$0.10 and $[\rm{Fe/H}]$=-0.3$\pm$0.2, 
somewhat higher than $[\rm{Fe/H}]$=-0.56 from the integrated light spectroscopic database of \citet{AZ88}.  Using $VI$ photometry, \citet{ortolanihp1vi} give $R_{\odot}$=6.85$\pm$0.2 kpc and estimate $[\rm{Fe/H}]$$\sim$-1.5 based on a direct comparison to NGC 6752, pointing out the extensive blue HB and proximity to the Galactic center, suggesting that early high values of $[\rm{Fe/H}]$ were due to contamination by field stars.  Using near-IR photometry, \citet{davidge} performed a comparison to M 13, giving $E(B-V)$=0.74 and $(m-M)_{0}$=14.3, corresponding to $R_{\odot}$$\sim$7.25 kpc, while medium resolution $K$-band spectroscopy by \citet{sf04} led to $[\rm{Fe/H}]$=-1.30$\pm$0.09.  The compilation of near-IR bulge GGC photometry by \citet{v10} gives $R_{\odot}$=6.8 kpc and $[Fe/H]_{CG97}$=-1.12.  A proper motion-cleaned CMD was presented by \citet{ortolaniao}, which they used to estimate distances ranging from 6.8$\leq$$R_{\odot}$$\leq$7.5 kpc depending on methodology, and they claim that HP 1 may be one of the oldest GGCs based on its location in HB morphology-metallicity space.  High resolution optical spectra have been obtained by \citet{barbuyhp1a,barbuyhp1b}, who give $[\rm{Fe/H}]$=-1.06$\pm$0.10 based on eight radial velocity members, in good agreement with $[\rm{Fe/H}]$=-1.17$\pm$0.07 from low-resolution spectra of eight members by \citet{brunofors2}, who also claim that HP 1 is a likely candidate to
be the oldest bulge GGC.  We confirm the extended blue HB described by \citet{ortolaniao}, the most clearly visible among our sample, which is apparent in all three of the color-magnitude planes shown in Fig.~\ref{cmdHP1} and has a tail extending from $F606W$$\sim$18.5 to at least $F606W$$\sim$20.5.

\subsection{NGC 6540 = Djorg 3}

\begin{figure}
\plotone{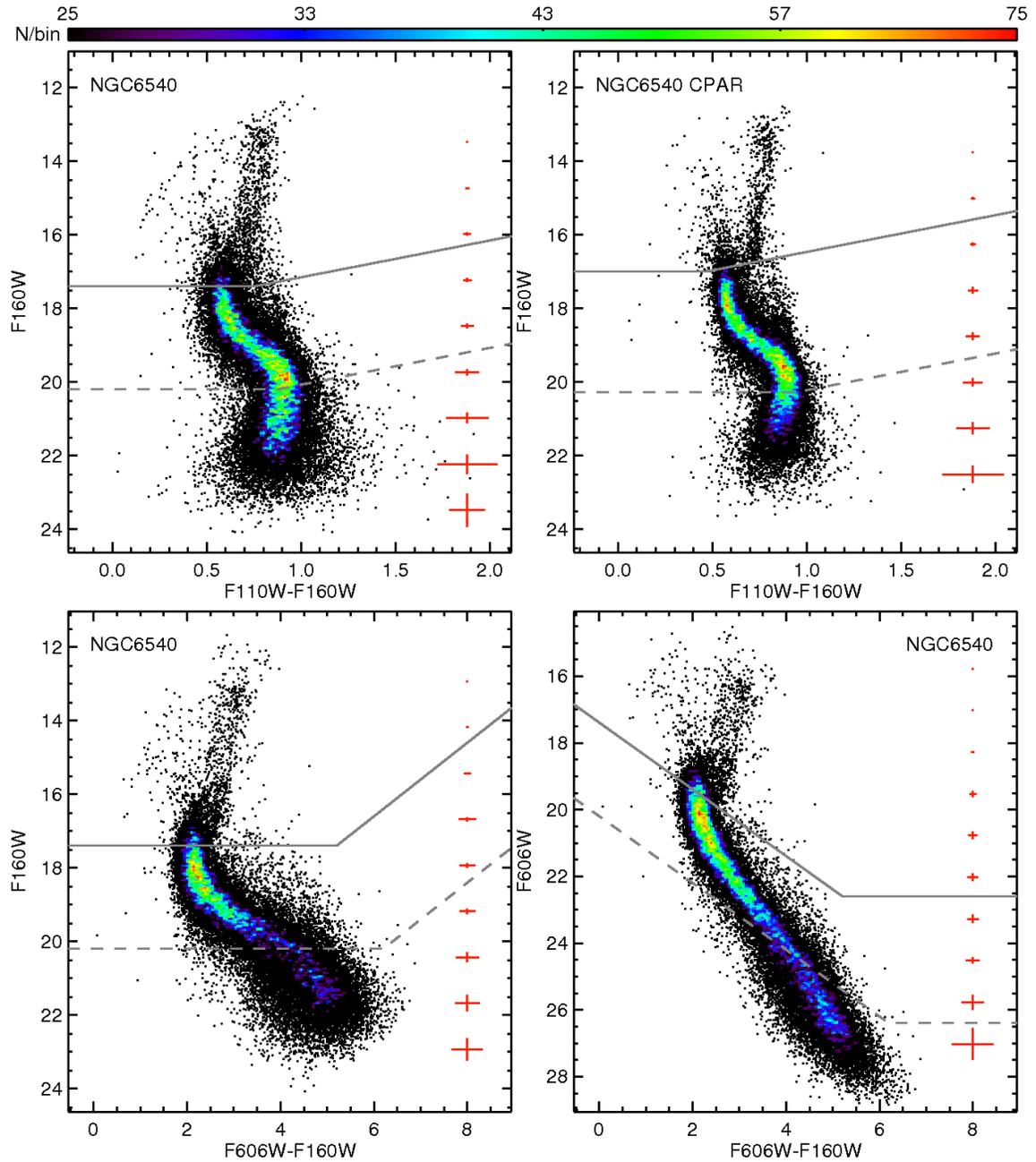}
\caption{As in Fig.~\ref{cmdfig1}, but for NGC 6540.
\label{cmdNGC6540}}
\end{figure}

Average values of $E(B-V)$=0.59$\pm$0.11 and $R_{\odot}$=3.5$\pm$0.4 kpc were estimated by \citet{bica6540}, who noted the blue horizontal branch and, based on optical photometry and integrated spectroscopy, claimed that this cluster is very concentrated and possibly post core collapse.  Using optical echelle spectra of six stars, \citet{cote} estimated $[\rm{Fe/H}]$=-1.4 and a mean radial velocity of -17.7$\pm$1.4 km/s.  This cluster was observed in the \citet{piottosnap} WFPC2 snapshot program, and the next photometric study was the near-infrared survey of bulge GCs by \citet{v10}, who also noted the sparseness of the cluster RGB and the concentrated nature of the cluster.   Their photometric calibrations and observed values of the RGB slope and RGB tip led to values of $[Fe/H]_{CG97}$=-1.29, $E(B-V)$=0.66 and $R_{\odot}$=5.2 kpc.  Proper motions of NGC 6540 are included in \citet{pmrossi}, who suggest that the cluster may be near apogalacticon based on the combination of low observed radial velocity and small proper motions which they measure relative to the foreground population.  However, they also point out the difficulty of proper motion cleaning in this particular case, and list a shorter adopted distance of $R_{\odot}$=3.7$\pm$0.3 kpc.  We confirm the blue HB extending to $F160W$$\sim$16 in Fig.~\ref{cmdNGC6540}, although the $(F606W,F606W-F160W)$ CMD illustrates that it lacks the extended tail seen in HP1.  Comparison of the primary and parallel field CMDs also suggests that, as with BH261, this cluster is relatively sparsely populated compared to the remainder of the target GGCs.

\subsection{NGC 6569}

\begin{figure}
\plotone{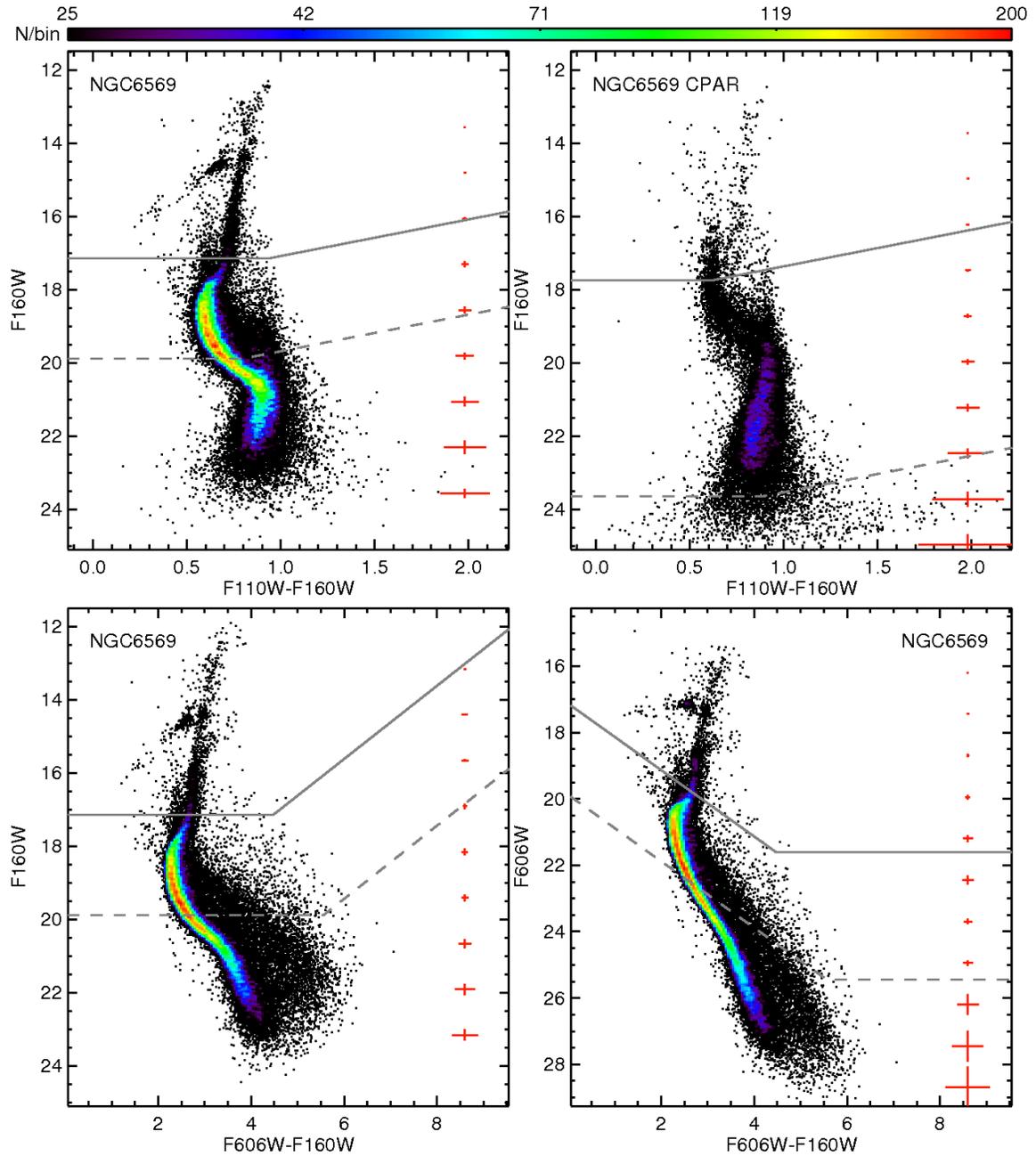}
\caption{As in Fig.~\ref{cmdfig1}, but for NGC 6569.
\label{cmdNGC6569}}
\end{figure}

Optical $BVI$ photometry was presented by \citet{ortolani6569}, who estimated $E(B-V)$=0.53 and $R_{\odot}$=9.8$\pm$1.0 kpc based on a comparison to 47 Tuc.  The compilation of \citet{v10} lists $E(B-V)$=0.49 and a longer distance of $R_{\odot}$=12.0 kpc, and \citet{6569spec} found $[\rm{Fe/H}]$=-0.79$\pm$0.02 based on high resolution near-IR spectroscopy of six cluster giants.  While optical photometry was available from the \citet{piottosnap} catalog, a double horizontal branch red clump was reported for the first time by \citet{maurohb} based on subsequent deeper near-IR photometry from the VVV survey.  The metallicity of NGC 6569 remains somewhat controversial, and this cluster is a 3$\sigma$ outlier in the \citet{maurocat} calibration of CaII triplet reduced equivalent width versus metallicity.  The CMD morphology in the near-IR similarly suggests a lower metallicity than the spectroscopic value ($[\rm{Fe/H}]$=-1.00$\pm$0.05; \citealt{cohenvvv}).  At the same time, \citet{brunofors2} give $[\rm{Fe/H}]$=-0.66$\pm$0.07 based on low resolution optical spectroscopy of seven members.  Recent high-resolution spectroscopy of a larger sample of cluster members by \citet{6569spec_johnson} found $[\rm{Fe/H}]$=-0.87 and evidence against metallicity or light-element variations as the cause of the double HB.  Meanwhile, cluster RR Lyrae variables have been recognized at least since \citet{6569oldvar}, and a detailed variability analysis by \citet{kunder6569} found that there was no evidence for helium enhancement among the cluster RR Lyrae.  As in the case of FSR1735, comparison of the primary and parallel fields in Fig.~\ref{cmdNGC6569} suggests that field contamination towards this cluster is fairly minimal given its distance from the Galactic plane.  In addition, the HB is seen to extend significantly blueward of the HB red clump, although the optical-IR CMDs do not reveal a blue HB extention as dramatic as seen in HP 1 or even NGC 6638.  

\subsection{NGC 6638}

\begin{figure}
\plotone{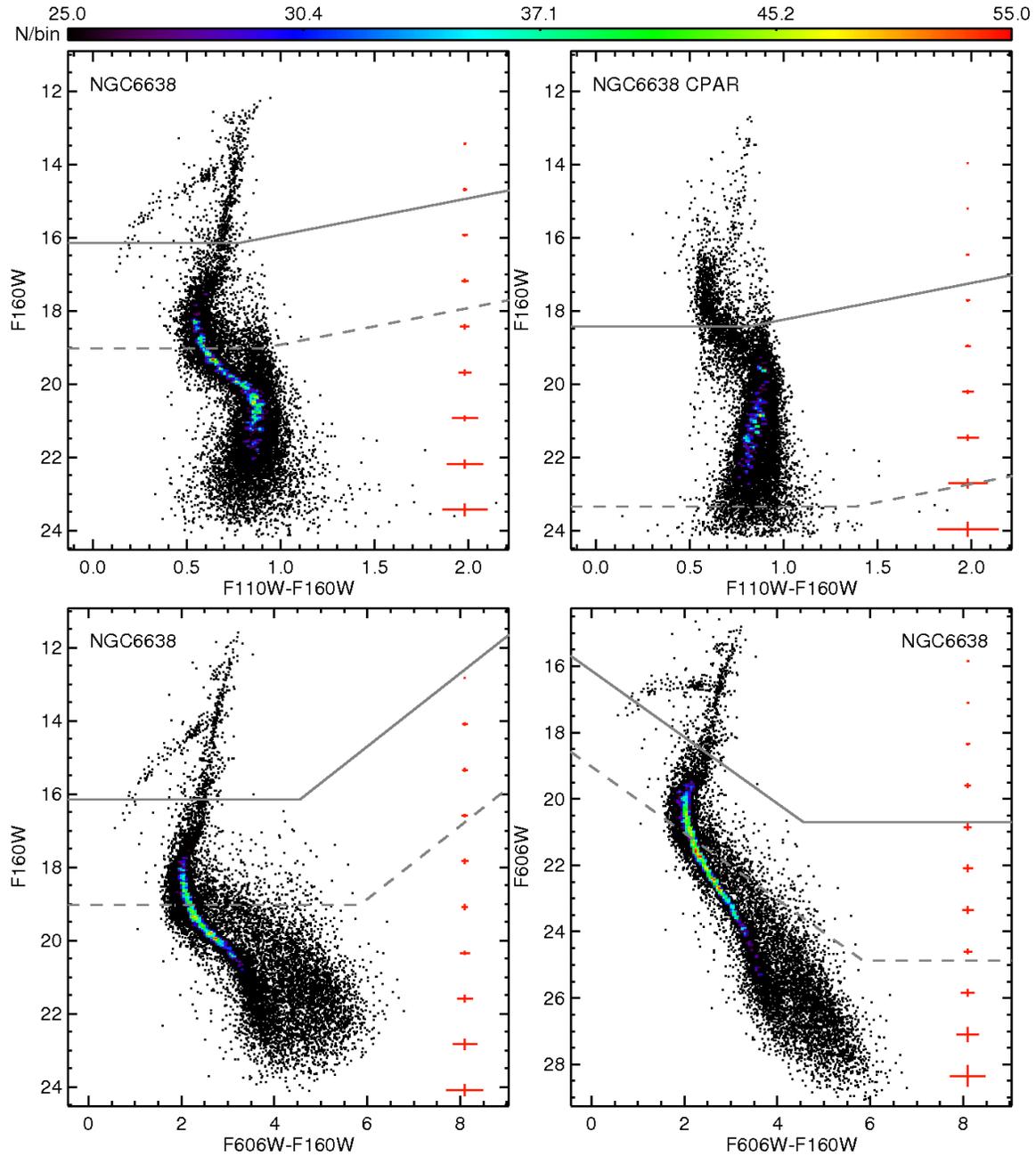}
\caption{As in Fig.~\ref{cmdfig1}, but for NGC 6638.
\label{cmdNGC6638}}
\end{figure}

A cluster CMD was presented by \citet{alcaino6638}, although they cautioned that their distance and reddening estimates should be viewed as tentative.  Low-resolution spectroscopy by \citet{rrlspec6638} of several RR Lyrae discovered by \citet{origvars6638}  yielded a metallicity estimate of $[\rm{Fe/H}]$=-0.82$\pm$0.2.  While the CaII triplet-based value of $[\rm{Fe/H}]$=-0.95$\pm$0.12 \citep{maurocat} is in reasonable agreement with the earlier estimate of $[\rm{Fe/H}]$=-0.99$\pm$0.07 \citep{c09} and the subsequent photometric estimate of $[\rm{Fe/H}]$=-1.09$\pm$0.07 \citep{cohenvvv}, there have been no high resolution spectroscopic studies of cluster members.  
As in the previous two cases, this cluster was included in the \citet{piottosnap} survey due to its relatively low foreground extinction.  Near-IR photometry by \citet{valenti05} led to estimates of $[Fe/H]_{CG97}$=-0.97, $E(B-V)$=0.43 and $R_{\odot}$=10.3 kpc. 
Along with HP1, this cluster presents the most clear evidence of a blue HB (see Fig.~\ref{cmdNGC6638}).  However, in this case the blue HB does not have an extensive tail although a small degree of "turnover" beyond the horizontal portion of the HB is seen in the $(F606W,F606W-F160W)$ CMD at $(F606W-F160W)$$\sim$1.1, $F606W$$\sim$16.5.  This cluster, along with NGC 6569, is among the least affected by differential reddening in our sample.        

\subsection{Palomar 6 = ESO 520-21}

\begin{figure}
\plotone{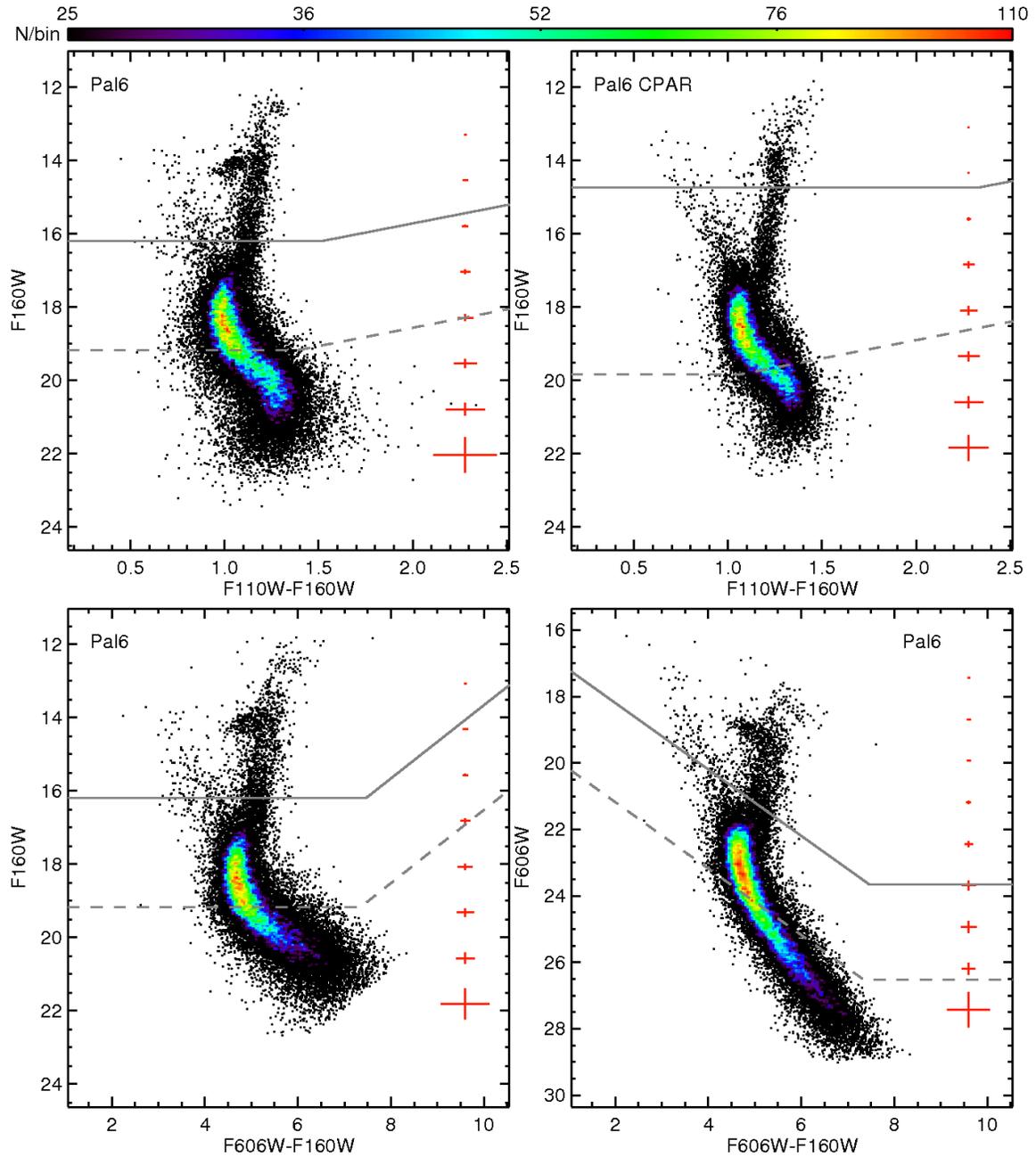}
\caption{As in Fig.~\ref{cmdfig1}, but for Palomar 6.
\label{cmdPal6}}
\end{figure}

Based on the locus of the HB and RGB in the $VI$ CMD, \citet{pal6vi} estimate $E(B-V)$=1.33 and $R_{\odot}$=8.95 kpc assuming $[\rm{Fe/H}]$$\sim$-0.4.  Using near-IR photometry and spectroscopy, \citet{minniti95b} list $E(J-K)$=0.91 and a super-solar metallicity of $[\rm{Fe/H}]$=0.2$\pm$0.3 based on measurements of three giants which they consider to be likely members.  Subsequently, near-IR photometry was presented by \citet{leepal6}, who estimated $E(B-V)$=1.30 and a distance modulus corresponding to $R_{\odot}$=7.2 kpc.  They also found that their measured value of the RGB slope implies a metallicity of $[\rm{Fe/H}]$=-1.2, significantly lower than previous studies.  Near-IR echelle spectroscopy of seven stars \citep{leepal6spec} further constrained their metallicity estimate to $[\rm{Fe/H}]$=-1.0$\pm$0.1, although near-IR spectroscopy of five members by \citet{sf04} led to a much higher value of $[\rm{Fe/H}]$=-0.52$\pm$0.2.  Most recently, a value of $[\rm{Fe/H}]$=-0.85$\pm$0.11 was reported from low-resolution optical spectra of four stars considered to be radial velocity members \citep{brunofors2}, although the $[\rm{Fe/H}]$ values which they derive based on empirical versus synthetic spectral libraries differ by more than 1 dex.  The red horizontal branch was confirmed by \citet{pmrossi}, who present cleaned CMDs and proper motion measurements.  The red HB is also clearly visible in Fig.~\ref{cmdPal6}, although differential reddening is particularly discernible from the optical-IR CMDs.  

\subsection{Terzan 1 = ESO 455-23}

\begin{figure}
\plotone{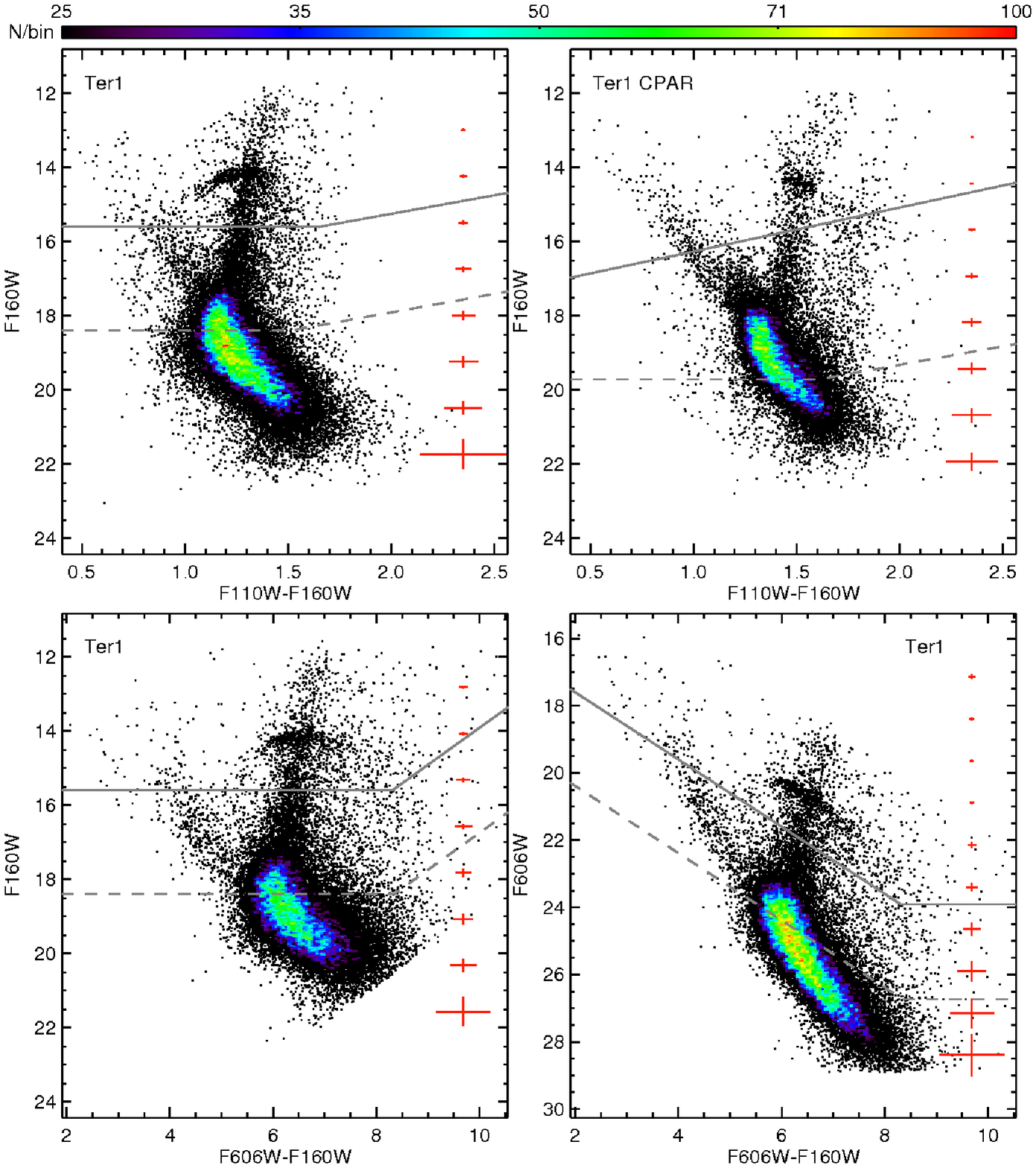}
\caption{As in Fig.~\ref{cmdfig1}, but for Terzan 1.
\label{cmdTer1}}
\end{figure}

Based on optical photometry, \citet{ortolaniter1orig} estimated a distance of $R_{\odot}$=4-5 kpc, and noted the differences between the field population close to Terzan 1 and that of Baade's Window.  In a subsequent study using \textit{HST} WFPC2 optical imaging, \citet{ortolaniter1} estimate $E(B-V)$=5.2$\pm$0.5 and $R_{\odot}$=5.2$\pm$0.5 kpc, cautioning that this distance depends strongly on the adopted reddening law.  They also note that the metallicity inferred from the RGB morphology ($[\rm{Fe/H}]$$\sim$-1.2) is at odds with the relatively red HB morphology.  Optical spectroscopy of 11 velocity members by \citet{idiart} yielded $[\rm{Fe/H}]$=-1.27$\pm$0.05.  Most recently, high resolution near-IR spectroscopy of 15 members by \citet{valentiter1} yielded values of $[\rm{Fe/H}]$=-1.26$\pm$0.03 and high ($\sim$0.4) $\alpha$-enhancement, while \citet{v10} report $E(B-V)$=1.99 and $R_{\odot}$=6.6 kpc based on a comparison of near-IR photometry to 47 Tuc.  Using two epochs of ground-based photometry, \citet{pmrossi} claim a very small relative proper motion, suggesting that in combination with the radial velocity reported by \citet{idiart}, the cluster could be on an elongated orbit, although \citet{valentiter1} find a significantly smaller (by a factor of nearly 2) radial velocity.  Lastly, it bears mention that x-ray emission has been detected from this cluster \citep{ter1x1,ter1x2,ter1x3,ter1x4,ter1x5,ter1x6}, including several sources which are likely located within the cluster along the line of sight \citep{ter1x7}.  From the IR CMDs of the primary and parallel fields in the top row of Fig.~\ref{cmdTer1}, it appears that there is a shift in the foreground/background RGB, although a density profile will clarify to what extent the cluster contributes to the reddest RGB stars seen in the target field.  Also, the HB may extend significantly blueward of the clump seen at $(F110W-F160W)$=1.2, possibly as far as $(F110W-F160W)$$\sim$0.7 at $F160W$$\sim$16, but careful field decontamination is needed to clarify the HB morphology of this cluster. 

\subsection{Terzan 2 = ESO 454-29}  

\begin{figure}
\plotone{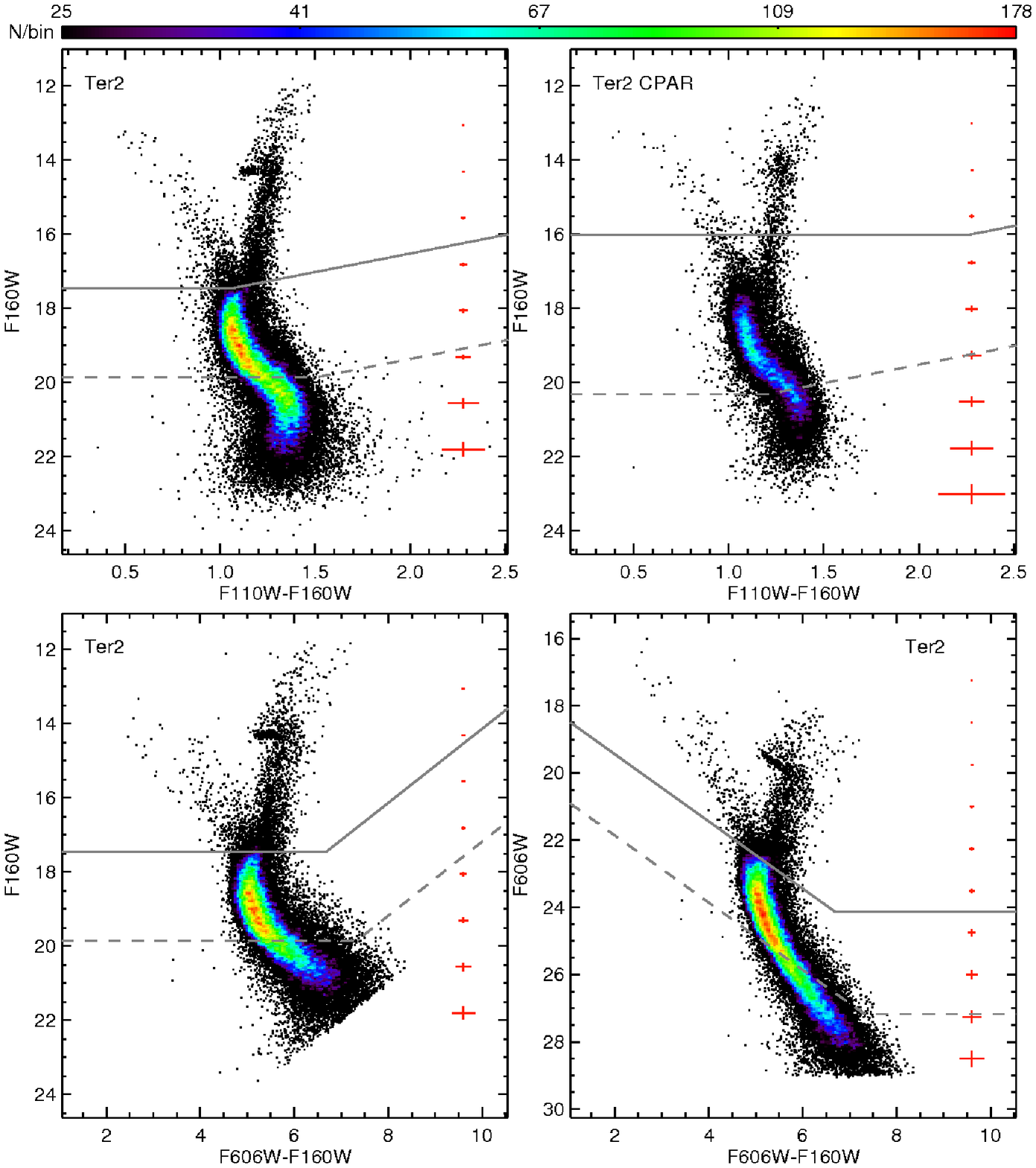}
\caption{As in Fig.~\ref{cmdfig1}, but for Terzan 2.
\label{cmdTer2}}
\end{figure}

Near-IR photometry was used by \citet{ter2ir} to estimate $R_{\odot}$$\sim$10 kpc and $E(B-V)$=1.25$\pm$0.15 via comparison to then-available template GGC giant branches, assuming $[\rm{Fe/H}]$=-0.25$\pm$0.15 from \citet{AZ88}.  The first optical CMD was presented by \citet{ortolaniter2}, who estimate $E(B-V)$=1.57 and $R_{\odot}$=7.7 kpc assuming a standard reddening law (or shorter distances for non-standard values of $R_{V}$$>$3.1).
Based on spectra of seven stars, \citet{sf04} estimate $[\rm{Fe/H}]$=-0.87$\pm$0.06, and \citet{v10} use their near-IR photometry to estimate $E(B-V)$=1.87, $R_{\odot}$=7.4 kpc and $[Fe/H]_{CG97}$=-0.72.  A low proper motion of $\lesssim$1 mas/yr is given by \citet{pmrossi} based on two epochs of ground-based optical imaging, and there have been several studies of x-ray sources located towards this cluster \citep[e.g.][]{ter2x1,ter2x2,ter2x3,ter2x4,ter2x5,ter2x6}.  A 
short red HB is clearly visible in our CMDs in Fig.~\ref{cmdTer2}, although the optical-IR CMDs reveal significant differential reddening.  

\subsection{Terzan 4 = Gcl-66.1}

\begin{figure}
\plotone{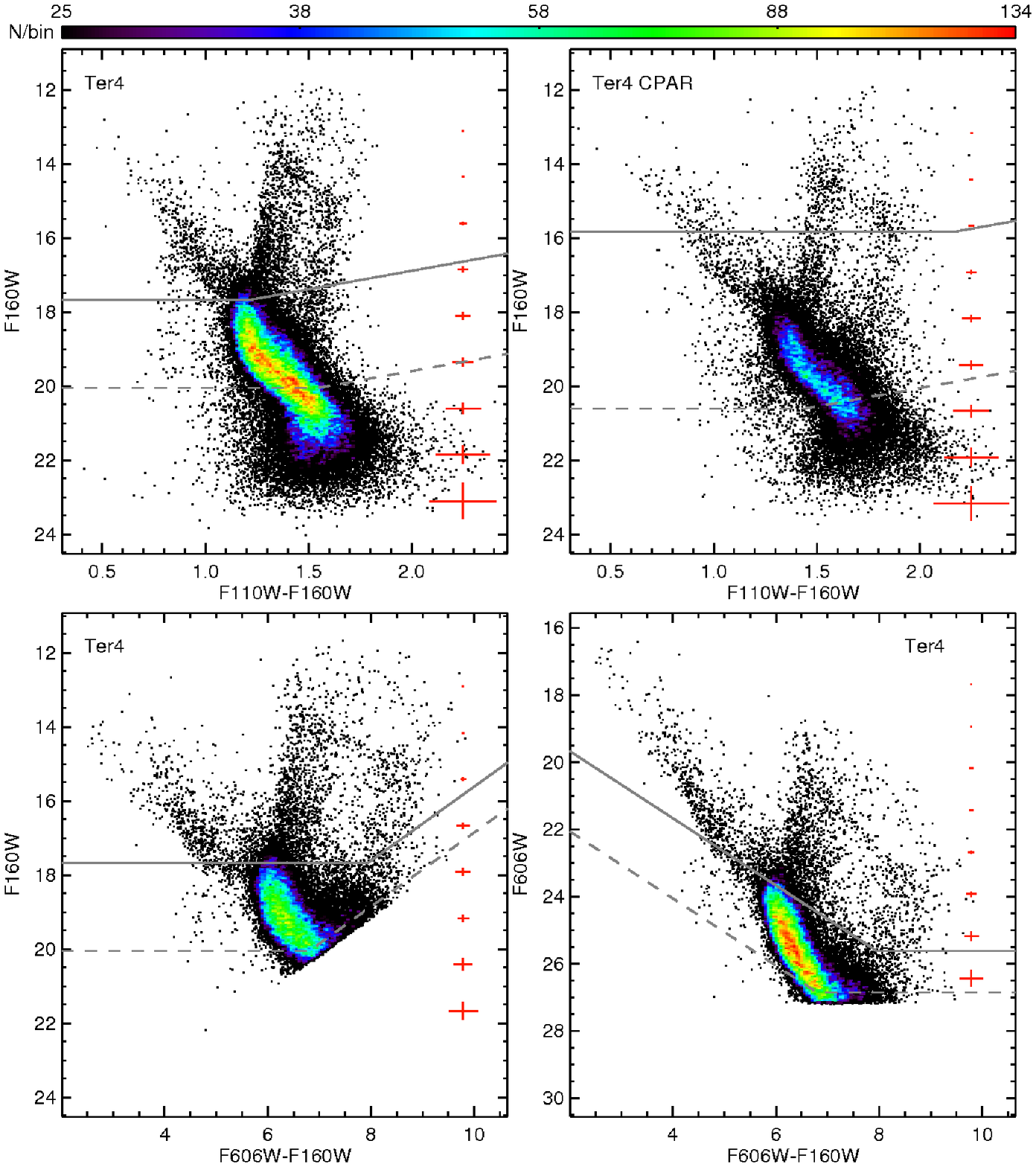}
\caption{As in Fig.~\ref{cmdfig1}, but for Terzan 4.
\label{cmdTer4}}
\end{figure}

\citet{ortolaniter4} estimate $E(B-V)$=2.35 and $R_{\odot}$=8.3 kpc, noting the blue horizontal branch of this cluster, and subsequent \textit{HST} NICMOS observations \citep{nicmos1} are used to calculate a distance of $R_{\odot}$=8.0$\pm$0.3 kpc via direct comparison with the GGC M 92 \citep{nicmos2}.  Using their database of near-IR photometry, \citet{v10} report $E(B-V)$=2.05 and $R_{\odot}$=6.7 kpc.  Medium resolution near-IR spectroscopy gives $[\rm{Fe/H}]$=-1.62$\pm$0.08 \citep{sf04}, in excellent agreement with the value of $[\rm{Fe/H}]$$\sim$-1.6 from high resolution IR spectra of four members by \citet{or04}, who also report a high value of $[\alpha/Fe]$$\sim$0.5.  Proper motions from \citet{pmrossi} are relatively large compared to the majority of their sample of GGCs towards the bulge, and they suggest that this, in combination with its relatively low metallicity, implies a possible halo origin for this cluster.  A very low concentration parameter of $c$=0.9$\pm$0.2 was inferred from fits of King profiles to 2MASS density and surface brightness profiles \citep{sb2mass}.  Our CMDs in Fig.~\ref{cmdTer4} illustrate severe differential reddening across the primary field, and as in the case of Terzan 1, a clear change in the location of the bulge RGB between the primary and parallel fields.  

\subsection{Terzan 6 = ESO 455-49}

\begin{figure}
\plotone{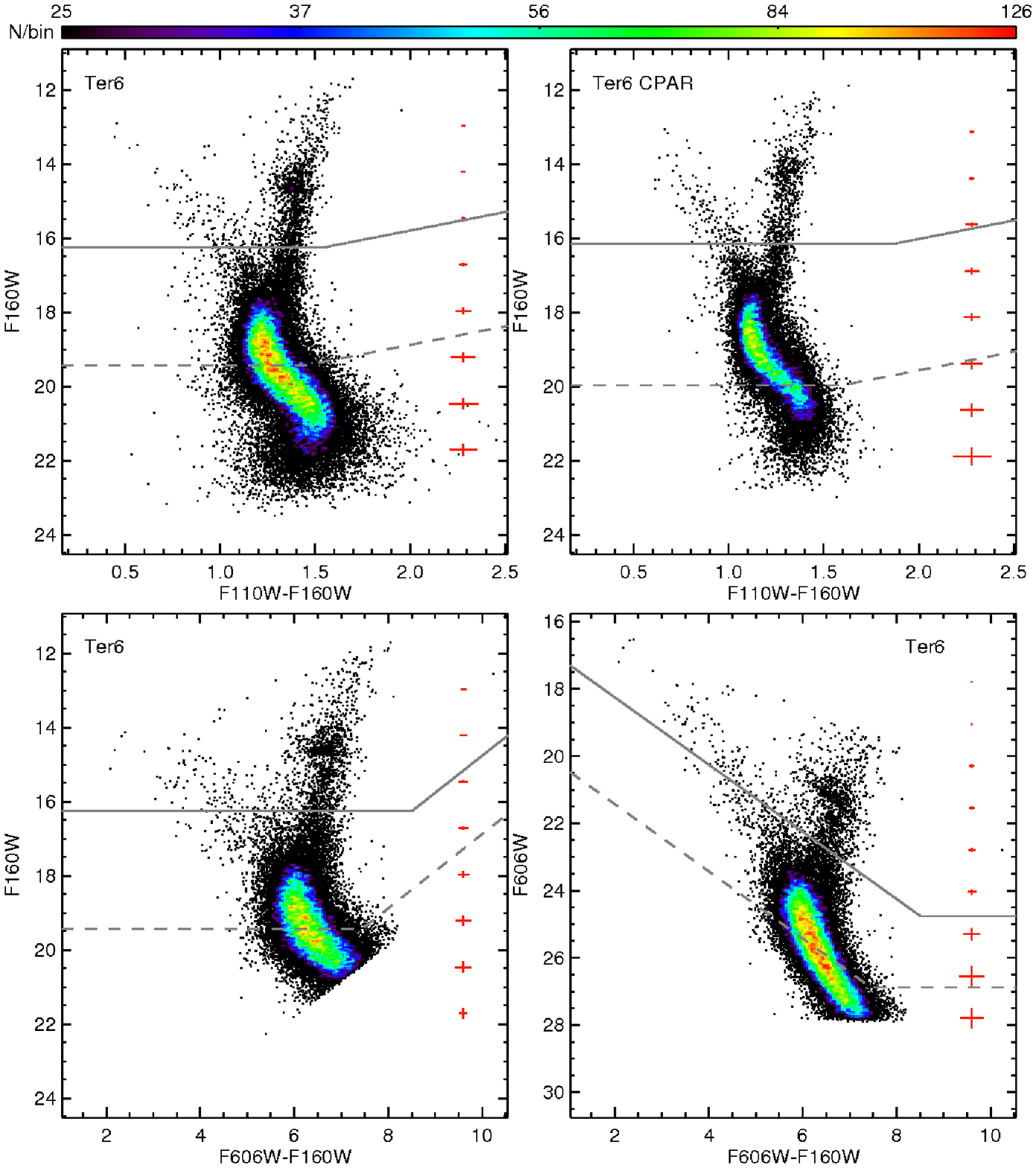}
\caption{As in Fig.~\ref{cmdfig1}, but for Terzan 6.
\label{cmdTer6}}
\end{figure}

Near-IR photometry was presented by \citet{ter6ir}, who estimated $E(B-V)$=2.04 and $R_{\odot}$=6.8$\pm$0.46 kpc based on a direct comparison to the GGC M 71.  Shortly afterward, \citet{ortolaniter6} use an optical CMD to estimate similar values of $E(B-V)$=2.24 and $R_{\odot}$=7.0 kpc assuming a standard reddening law.  The only subsequent photometric study is the near-IR database of \citet{v10}, estimating $E(B-V)$=2.35 and $R_{\odot}$=6.7 kpc.  There is one known low-mass X-ray binary in Terzan 6 \citep[e.g.][]{ter6x1,ter6x2}.  As in the case of Terzan 2, our CMDs confirm a short red HB, while the reddening-sensitive optical-IR CMDs in the bottom row of Fig.~\ref{cmdTer6} reveal significant differential reddening across the target field.

\subsection{Terzan 9 = Gcl-80.1}

\begin{figure}
\plotone{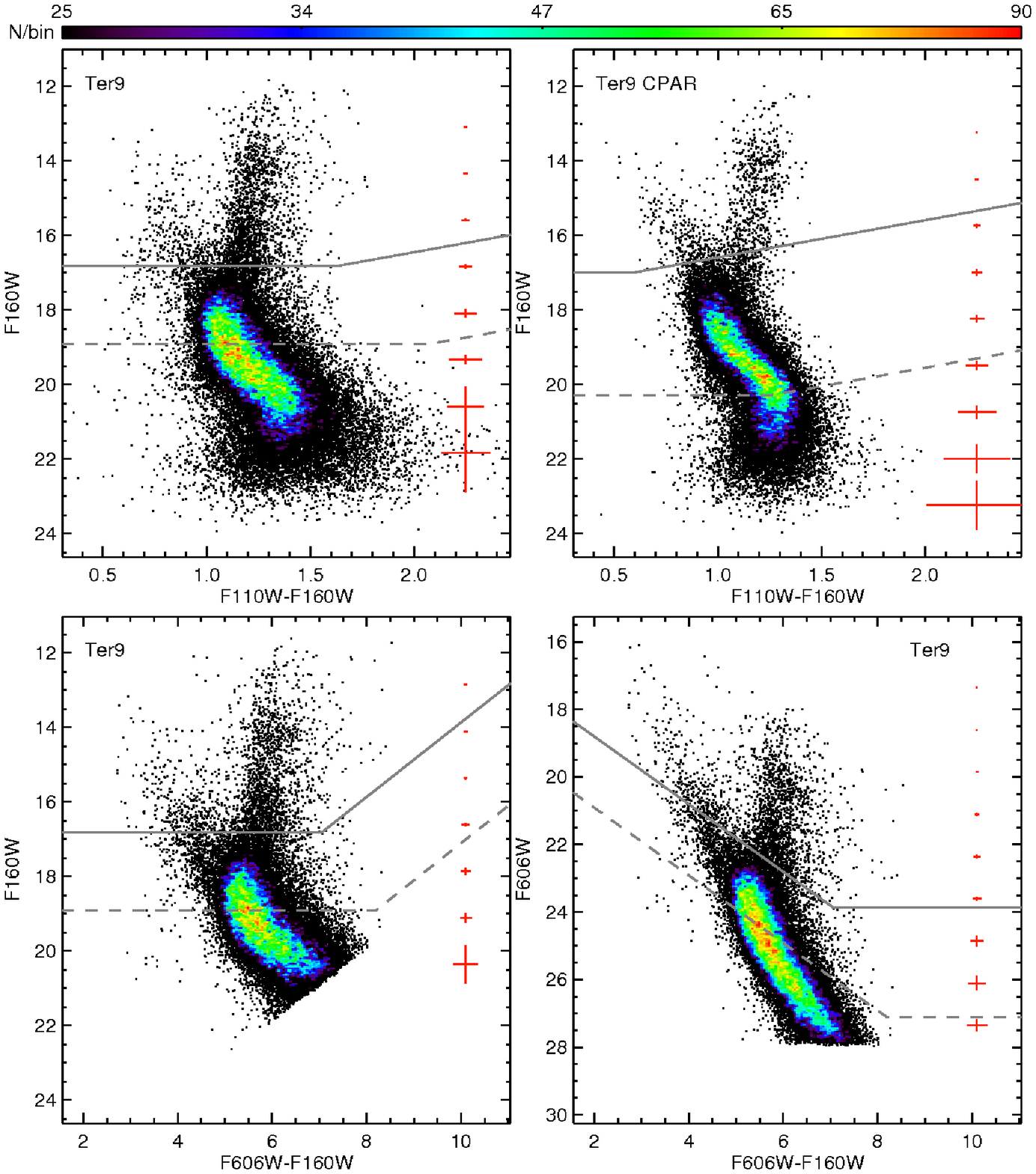}
\caption{As in Fig.~\ref{cmdfig1}, but for Terzan 9.
\label{cmdTer9}}
\end{figure}

\begin{figure}
\plotone{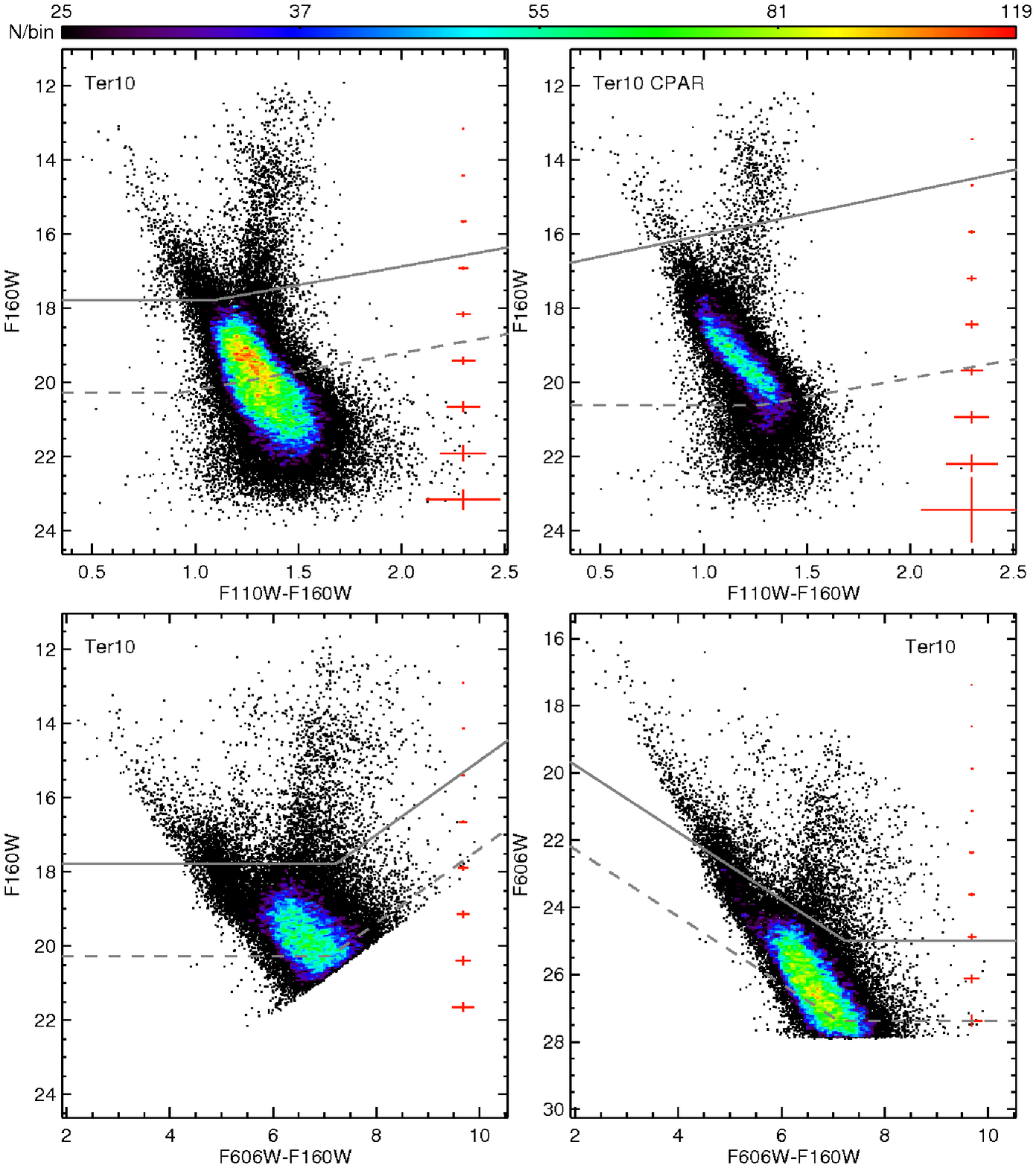}
\caption{As in Fig.~\ref{cmdfig1}, but for Terzan 10.
\label{cmdTer10}}
\end{figure}

The first optical CMD was presented by \citet{ortolaniter9}, who estimate $E(B-V)$=1.95 and $R_{\odot}$=4.9$\pm$0.7 kpc via direct comparison to the GGC M 30.  Using near-IR photometry, \citet{v10} use the measured RGB slope and tip to obtain $E(B-V)$=1.79 and $R_{\odot}$=5.6 kpc.  The only spectroscopic metallicity estimate remains the CaII triplet study of \citet{AZ88}, giving $[\rm{Fe/H}]$=-0.99.  Using a second epoch of optical imaging, \citet{pmrossi} mention that this cluster has a relatively high proper motion compared to other bulge GGCs.  Our CMDs in Fig.~\ref{cmdTer9} suggest that this cluster is both fairly sparse and severely affected by differential reddening, so any morphological conclusions must be postponed to a future study presenting differential reddening corrected CMDs.

\subsection{Terzan 10 = ESO 521-16}

An optical CMD was presented by \citet{djorg2discov}, but they find that the cluster HB is at the detection limit of their data.  They note the large differential reddening towards this cluster and estimate $R_{\odot}$=4.8$\pm$1.0 kpc.  This value was substantially revised by \citet{javiergcrrl} using near-IR time series imaging from VVV, and by exploiting the differential reddening among cluster RRL, they find an unusually flat near-IR extinction law of $A_{Ks}/E(J-K_{S})$=0.47$\pm$0.05, yielding a distance of $R_{\odot}$=10.3$\pm$0.2$\pm$0.2 kpc, which places Terzan 10 on the opposite (far) side of the Galactic bulge from previous studies.  This is another cluster with a low concentration parameter according to \citet{sb2mass}, who report $c$=0.8$\pm$0.2 or 0.7$\pm$0.3 based on King fits to 2MASS density profiles and surface brightness profiles respectively.  Despite our smaller ($\sim$200$\arcsec$ per side) field of view as compared to \citet{javiergcrrl}, our photometry also illustrates severe differential reddening across the target field of view.  

\subsection{Terzan 12 = ESO 522-1}

\begin{figure}
\plotone{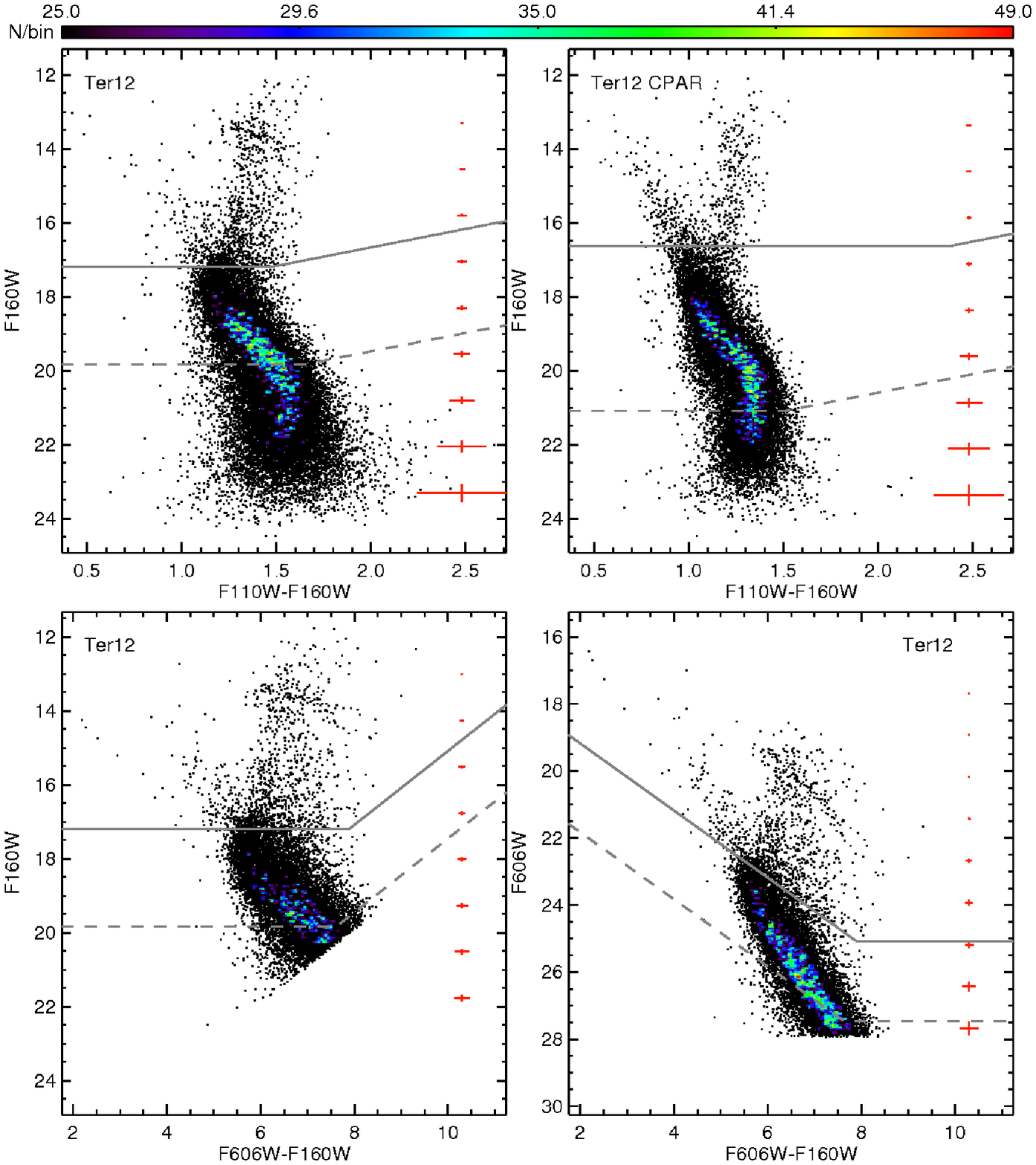}
\caption{As in Fig.~\ref{cmdfig1}, but for Terzan 12.
\label{cmdTer12}}
\end{figure}

The optical CMD presented by \citet{ortolaniter12} is the only published one for this cluster, and they use the HB to calculate $E(B-V)$=2.06$\pm$0.06 and $R_{\odot}$=3.4$\pm$0.5 kpc.  Based on high resolution optical spectroscopy of three stars, \citet{cote} list $[\rm{Fe/H}]$=-0.5 and a radial velocity of 94.1$\pm$1.5 km/s.  Our photometry in Fig.~\ref{cmdTer12} shows significant differential reddening, and also a likely red HB.   

\subsection{Ton 2 = Pismis 26 = ESO 333-16}

\begin{figure}
\plotone{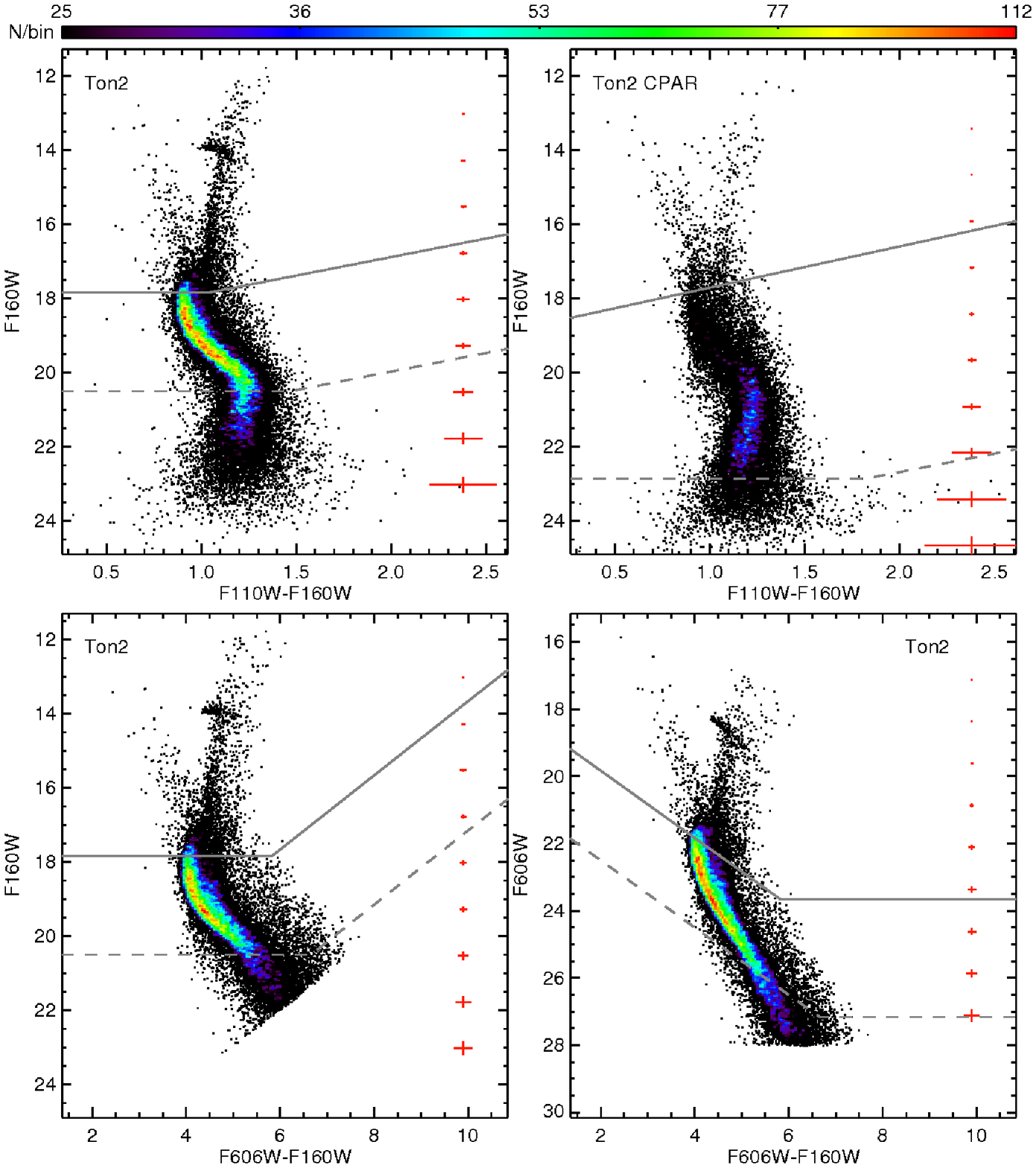}
\caption{As in Fig.~\ref{cmdfig1}, but for Ton 2. \label{cmdfiglast}}
\end{figure}

An optical $VI$ CMD was presented by \citet{ortolaniton2}, who use a direct comparison with 47 Tuc to estimate $E(B-V)$=1.26 and $R_{\odot}$=6.4$\pm$0.6 kpc.  The only spectroscopic abundance value is from \citet{cote}, who finds $[\rm{Fe/H}]$=-0.6 and a radial velocity of -184.4$\pm$2.2 km/s from optical echelle spectra of three stars.  The cluster CMDs in Fig.~\ref{cmdfiglast} reveal low field contamination and a clear red HB.  However, the shape of the red HB in the optical-IR CMDs clarifies that despite the clear definition of the cluster sequences in this case, differential reddening persists over the cluster field.

\section{Summary}

We have presented the observation and data reduction strategy for multi-wavelength space-based imaging of 16 GGCs located towards the inner Galactic bulge and disk.  Even a cursory glance at the cluster and parallel field CMDs (uncorrected for differential reddening) indicates the diversity of cluster properties (metallicity, density, HB morphology) sampled by our target clusters.  However, these CMDs also reveal a remarkable variation in the distribution of bulge and disk populations along the various sightlines probed by our imaging, useful in their own right to constrain models of Galactic structure and extinction towards the inner Milky Way.  Forthcoming studies will exploit this imaging to quantitatively address and refine existing values (based largely on optical imaging) of cluster structural parameters, distances and HB morphology, using differential reddening maps to optimize our photometry, while measuring target cluster ages, mass functions and mass segregation for the first time.  Lastly, due to its spatial resolution, our imaging comprises a first epoch enabling future proper motion studies of our target clusters as well as their foreground and background disk and bulge components.

The HST imaging used in our analysis is available at \dataset[10.17909/T9HD59]{https://doi.org/10.17909/T9HD59}.

\acknowledgements

It is a pleasure to thank the anonymous referee for their valuable comments and suggestions.   
R.E.C.~acknowledges support from Gemini-CONICYT for Project 32140007.  J.A-G.~acknowledges support by FONDECYT Iniciaci\'on 11150916, MINEDUC ANT 1655 and by the Ministry of Economy, Development, and Tourism’s Millennium Science Initiative through grant IC120009, awarded to the Millennium Institute of Astrophysics (MAS).  M.H.~acknowledges support from BASAL CATA Center for Astrophysics and Associated Technologies PFB-06.  D.G.~and F.M.~also acknowledge
financial support from the Chilean BASAL Centro de Excelencia en Astrofisica y Technologias Afines 
(CATA) grant PFB-06/2007.  This publication makes use of
observations collected at the European Organization for Astronomical Research in the Southern 
Hemisphere, Chile, under ESO program 179.B-2002 (VVV survey), as well as the SIMBAD databased, operated at CDS, Strasbourg, France.

\vspace{5mm}
\facilities{HST (ACS,WFC3), ESO:VISTA}



\begin{thebibliography}{}

\bibitem[Alcaino \& Liller(1983)]{alcaino6638} Alcaino, G., \& Liller, W.\ 1983, \aj, 88, 1166 

\bibitem[Alonso-Garc{\'{\i}}a et al.(2012)]{ag12} Alonso-Garc{\'{\i}}a, J., Mateo, M., Sen, B., et al.\ 2012, \aj, 143, 70 

\bibitem[Alonso-Garc{\'{\i}}a et al.(2015)]{javiergcrrl} Alonso-Garc{\'{\i}}a, J., D{\'e}k{\'a}ny, I., Catelan, M., et al.\ 2015, \aj, 149, 99

\bibitem[Altamirano et al.(2008)]{ter2x5} Altamirano, D., van der Klis, M., M{\'e}ndez, M., et al.\ 2008, \apj, 687, 488-504 

\bibitem[Anderson \& van der Marel(2010)]{omegacenpm} Anderson, J., \& van der Marel, R.~P.\ 2010, \apj, 710, 1032 

\bibitem[Armandroff \& Zinn(1988)]{AZ88} Armandroff, T.~E., \& Zinn, R.\ 1988, \aj, 96, 92 

\bibitem[Bahramian et al.(2016)]{ter6x2} Bahramian, A., Heinke, C.~O., Sivakoff, G.~R., et al.\ 2016, The Astronomer's Telegram, 9072, 1

\bibitem[Barbuy et al.(1997)]{ortolaniter6} Barbuy, B., Ortolani, S., \& Bica, E.\ 1997, \aaps, 122, 483

\bibitem[Barbuy et al.(2006)]{barbuyhp1a} Barbuy, B., Zoccali, M., Ortolani, S., et al.\ 2006, \aap, 449, 349 

\bibitem[Barbuy et al.(2016)]{barbuyhp1b} Barbuy, B., Cantelli, E., Vemado, A., et al.\ 2016, \aap, 591, A53 

\bibitem[Barret et al.(1991)]{ter2x1} Barret, D., Mereghetti, S., Roques, J.~P., et al.\ 1991, \apjl, 379, L21 

\bibitem[Barret et al.(1999)]{ter2x4} Barret, D., Grindlay, J.~E., Harrus, I.~M., \& Olive, J.~F.\ 1999, \aap, 341, 789

\bibitem[Bellini et al.(2014)]{hstpromo1} Bellini, A., Anderson, J., van der Marel, R.~P., et al.\ 2014, \apj, 797, 115

\bibitem[Bica et al.(1994)]{bica6540} Bica, E., Ortolani, S., \& Barbuy, B.\ 1994, \aap, 283, 67

\bibitem[Bica et al.(1996)]{ortolaniton2} Bica, E., Ortolani, S., \& Barbuy, B.\ 1996, \aaps, 120, 153 

\bibitem[Bica et al.(2016)]{bicarev} Bica, E., Ortolani, S., \& Barbuy, B.\ 2016, \pasa, 33, e028

\bibitem[Bohlin(2012)]{bohlin} Bohlin, R.~C.\ 2012, Instrument Science Report ACS 2012-01

\bibitem[Bonatto \& Bica(2008a)]{bbsb} Bonatto, C., \& Bica, E.\ 2008a, \aap, 477, 829 

\bibitem[Bonatto \& Bica(2008b)]{sb2mass} Bonatto, C., \& Bica, E.\ 2008b, \aap, 479, 741

\bibitem[Bonatto et al.(2009)]{fsr1767} Bonatto, C., Bica, E., Ortolani, S., \& Barbuy, B.\ 2009, \mnras, 397, 1032

\bibitem[Borrel et al.(1996)]{ter1x3} Borrel, V., Bouchet, L., Jourdain, E., et al.\ 1996, \aaps, 120, 249 

\bibitem[Brown et al.(2010)]{brownwfc3} Brown, T.~M., Sahu, K., Anderson, J., et al.\ 2010, \apjl, 725, L19

\bibitem[Cackett et al.(2006)]{ter1x7} Cackett, E.~M., Wijnands, R., Heinke, C.~O., et al.\ 2006, \mnras, 369, 407 

\bibitem[Carballo-Bello et al.(2016)]{fsr1735vvv} Carballo-Bello, J.~A., Ram{\'{\i}}rez Alegr{\'{\i}}a, S., Borissova, J., et al.\ 2016, \mnras, 462, 501

\bibitem[Carretta et al.(2009)]{c09} Carretta, E., Bragaglia, A., Gratton, R., D'Orazi, V., \& Lucatello, S.\ 2009, \aap, 508, 695 

\bibitem[Christian \& Friel(1992)]{ter2ir} Christian, C.~A., \& Friel, E.~D.\ 1992, \aj, 103, 142 

\bibitem[Cohen \& Sarajedini(2012)]{cohensxphe} Cohen, R.~E., \& Sarajedini, A.\ 2012, \mnras, 419, 342 

\bibitem[Cohen at al.(2014)]{cohen6544} Cohen, R.~E., Mauro, F., Geisler, D., et al.\ 2014, \aj, 148, 18

\bibitem[Cohen at al.(2015)]{cohenispi}  Cohen, R. E., Hempel, M., Mauro, F., et al. \ 2015, \aj, 150, 176

\bibitem[Cohen at al.(2017)]{cohenvvv}  Cohen, R.~E., Moni Bidin, C., Mauro, F., Bonatto, C., \& Geisler, D.\ 2017, \mnras, 464, 1874 

\bibitem[Correnti et al.(2016)]{matteoir} Correnti, M., Gennaro, M., Kalirai, J.~S., Brown, T.~M., \& Calamida, A.\ 2016, \apj, 823, 18

\bibitem[C{\^o}t{\'e}(1999)]{cote} C{\^o}t{\'e}, P.\ 1999, \aj, 118, 406 

\bibitem[Dalessandro et al.(2016)]{gd6624} Dalessandro, E., Saracino, S., Origlia, L.,  et al.\ 2016, \apj, 833, 111

\bibitem[Davidge(2000)]{davidge} Davidge, T.~J.\ 2000, \aj, 120, 1853 

\bibitem[De Angeli et al.(2005)]{deangeli} De Angeli, F., Piotto, G., Cassisi, S., et al.\ 2005, \aj, 130, 116 

\bibitem[Dias et al.(2016)]{brunofors2} Dias, B., Barbuy, B., Saviane, I., et al.\ 2016, \aap, 590, A9 

\bibitem[Dieball et al.(2016)]{m4bdir} Dieball, A., Bedin, L.~R., Knigge, C., et al.\ 2016, \apj, 817, 48 

\bibitem[Dolphin(2000)]{dolphin} Dolphin, A.~E.\ 2000, PASP, 112, 1383

\bibitem[Dotter et al.(2010)]{d10} Dotter, A., Sarajedini, A., Anderson, J., et al.\ 2010, \apj, 708, 698 

\bibitem[Dotter et al.(2011)]{dotteroh} Dotter, A., Sarajedini, A., \& Anderson, J.\ 2011, \apj, 738, 74 

\bibitem[Dotter et al.(2015)]{dotter6752} Dotter, A., Ferguson, J.~W., Conroy, C., et al.\ 2015, \mnras, 446, 1641 

\bibitem[Dressel(2018)]{wfc3handbook} Dressel, L.\ 2018, "Wide Field Camera 3 Instrument Handbook, Version 10.0" (Baltimore:STScI)

\bibitem[Fahlman et al.(1995)]{ter6ir} Fahlman, G.~G., Douglas, K.~A., \& Thompson, I.~B.\ 1995, \aj, 110, 2189 

\bibitem[Fern{\'a}ndez-Trincado et al.(2018)]{jose6522} Fern{\'a}ndez-Trincado, J.~G., Zamora, O., Souto, D., et al.\ 2018, arXiv:1801.07136

\bibitem[Ferraro et al.(2006)]{ferraromethod} Ferraro, F.~R., Valenti, E., \& Origlia, L.\ 2006, \apj, 649, 243 

\bibitem[Ferraro et al.(2009)]{ter5hb} Ferraro, F.~R., Dalessandro, E., Mucciarelli, A., et al.\ 2009, \nat, 462, 483 

\bibitem[Ferraro et al.(2016)]{ter5age} Ferraro, F.~R., Massari, D., Dalessandro, E., et al.\ 2016, \apj, 828, 75 

\bibitem[Fiorentino et al.(2016)]{gemsrev} Fiorentino, G., Massari, D., McConnachie, A., et al.\ 2016, arXiv:1608.01457

\bibitem[Froebrich et al.(2007)]{fsr1735discov} Froebrich, D., Meusinger, H., \& Scholz, A.\ 2007, \mnras, 377, L54 

\bibitem[Goldsbury et al.(2010)]{goldsbury10} Goldsbury, R., Richer, H.~B., Anderson, J., et al.\ 2010, \aj, 140, 1830-1837

\bibitem[Goldsbury et al.(2013)]{goldsbury13} Goldsbury, R., Heyl, J., \& Richer, H.\ 2013, \apj, 778, 57 

\bibitem[Gonzalez et al.(2012)]{gonzalezmap} Gonzalez, O.~A., Rejkuba, M., Zoccali, M., et al.\ 2012, \aap, 543, A13

\bibitem[Guainazzi et al.(1998)]{ter2x3} Guainazzi, M., Parmar, A.~N., Segreto, A., et al.\ 1998, \aap, 339, 802 

\bibitem[Guainazzi et al.(1999)]{ter1x4} Guainazzi, M., Parmar, A.~N., \& Oosterbroek, T.\ 1999, \aap, 349, 819 

\bibitem[\protect\citeauthoryear{H96}{}]{h96} Harris, W.~E.\ 1996, \aj, 112, 1487 

\bibitem[Hesser et al.(1984)]{6569oldvar} Hesser, J.~E., Harris, H.~C., \& Harris, G.~L.~H.\ 1984, \baas, 16, 967 

\bibitem[Idiart et al.(2002)]{idiart} Idiart, T.~P., Barbuy, B., Perrin, M.-N., et al.\ 2002, \aap, 381, 472 

\bibitem[in't Zand et al.(2003)]{ter6x1} in't Zand, J.~J.~M., Hulleman, F., Markwardt, C.~B., et al.\ 2003, \aap, 406, 233 

\bibitem[Johnson et al.(2018)]{6569spec_johnson} Johnson, C.~I., Rich, R.~M., Caldwell, N., et al.\ 2018, \aj, 155, 71

\bibitem[Johnston et al.(1995)]{ter1x2} Johnston, H.~M., Verbunt, F., \& Hasinger, G.\ 1995, \aap, 298, L21 

\bibitem[Kerber et al.(2018)]{pm6522} Kerber, L.~O., Nardiello, D., Ortolani, S., et al.\ 2018, \apj, 853, 15 

\bibitem[King(1962)]{king62} King, I.\ 1962, \aj, 67, 471 

\bibitem[Kunder et al.(2015)]{kunder6569} Kunder, A., Stetson, P.~B., Catelan, M., et al.\ 2015, Fifty Years of Wide Field Studies in the Southern Hemisphere: Resolved Stellar Populations of the Galactic Bulge and Magellanic Clouds, 491, 104

\bibitem[LaGioia et al.(2014)]{6528hst} Lagioia, E.~P., Milone, A.~P., Stetson, P.~B., et al.\ 2014, \apj, 785, 81 

\bibitem[Lanzoni et al.(2010)]{ter5sb} Lanzoni, B., Ferraro, F.~R., Dalessandro, E., et al.\ 2010, \apj, 717, 653 

\bibitem[Leaman et al.(2013)]{leaman13} Leaman, R., VandenBerg, D.~A., \& Mendel, J.~T.\ 2013, \mnras, 436, 122 

\bibitem[Lee \& Carney(2002)]{leepal6} Lee, J.-W., \& Carney, B.~W.\ 2002, \aj, 123, 3305 

\bibitem[Lee et al.(2004)]{leepal6spec} Lee, J.-W., Carney, B.~W., \& Balachandran, S.~C.\ 2004, \aj, 128, 2388 

\bibitem[Makishima et al.(1981)]{ter1x1} Makishima, K., Ohashi, T., Inoue, H., et al.\ 1981, \apjl, 247, L23 

\bibitem[Mar{\'{\i}}n-Franch et al.(2009)]{mf09} Mar{\'{\i}}n-Franch, A., Aparicio, A., Piotto, G., et al.\ 2009, \apj, 694, 1498 

\bibitem[Mauro et al.(2012)]{maurohb} Mauro, F. et al., 2012, ApJL, 761, 29

\bibitem[Mauro et al.(2013)]{mauropipe} Mauro, F., Moni Bidin, C., Chen{\'e}, A.-N. et al., 2013, RMxAA, 49, 189

\bibitem[Mauro et al.(2014)]{maurocat} Mauro, F., Moni Bidin, C., Geisler, D., et al.\ 2014, \aap, 563, A76 

\bibitem[Massari et al.(2014)]{ter5feh2} Massari, D., Mucciarelli, A., Ferraro, F.~R., et al.\ 2014, \apj, 795, 22 

\bibitem[Massari et al.(2016)]{gems2808} Massari, D., Fiorentino, G., McConnachie, A., et al.\ 2016, \aap, 586, A51 

\bibitem[McLaughlin \& van der Marel (2005)]{mvdm} McLaughlin, D. E. \& van der Marel, R. P., 2005, ApJS, 161, 304

\bibitem[Mereghetti et al.(1995)]{ter2x2} Mereghetti, S., Barret, D., Stella, L., et al.\ 1995, \aap, 302, 713 

\bibitem[Milone et al.(2008)]{milone1851} Milone, A.~P., Bedin, L.~R., Piotto, G., et al.\ 2008, \apj, 673, 241-250 

\bibitem[Milone et al.(2012)]{milonebinfrac} Milone, A.~P., Piotto, G., Bedin, L.~R., et al.\ 2012, \aap, 540, A16

\bibitem[Milone et al.(2015)]{milone2808} Milone, A.~P., Marino, A.~F., Piotto, G., et al.\ 2015, \apj, 808, 51 

\bibitem[Minniti et al.(1995)]{minniti95a} Minniti, D., Olszewski, E.~W., \& Rieke, M.\ 1995, \aj, 110, 1686 

\bibitem[Minniti(1995)]{minniti95b} Minniti, D.\ 1995, \aap, 303, 468

\bibitem[Minniti et al.(2010)]{minnitivvv} Minniti, D., Lucas, P.~W., Emerson, J.~P., et al.\ 2010, \na, 15, 433

\bibitem[Minniti et al.(2011)]{minniticl001} Minniti, D., Hempel, M., Toledo, I., et al.\ 2011, \aap, 527, A81

\bibitem[Minniti et al.(2017)]{minnitinewgc} Minniti, D., Palma, T., D{\'e}k{\'a}ny, I., et al.\ 2017, arXiv:1703.02033

\bibitem[Miocchi et al.(2013)]{miocchisb} Miocchi, P., Lanzoni, B., Ferraro, F.~R., et al.\ 2013, \apj, 774, 151 

\bibitem[Molkov et al.(2001)]{ter1x5} Molkov, S.~V., Grebenev, S.~A., \& Lutovinov, A.~A.\ 2001, Astronomy Letters, 27, 363 

\bibitem[Mu\~{n}oz et al.(2017)]{cesar6440} Mu\~{n}oz, C., Villanova, S., Geisler, D., et al.\ 2017, \aap, submitted

\bibitem[Ness et al.(2016)]{apogeebulge} Ness, M., Zasowski, G., Johnson, J.~A., et al.\ 2016, \apj, 819, 2

\bibitem[Noyola \& Gebhardt(2006)]{noyolasb} Noyola, E., \& Gebhardt, K.\ 2006, \aj, 132, 447 

\bibitem[O'Malley et al.(2017)]{chaboyerage} O'Malley, E.~M., Gilligan, C., \& Chaboyer, B.\ 2017, arXiv:1703.01915 

\bibitem[Origlia \& Rich(2004)]{or04} Origlia, L., \& Rich, R.~M.\ 2004, \aj, 127, 3422 

\bibitem[Origlia et al.(2013)]{ter5feh1} Origlia, L., Massari, D., Rich, R.~M., et al.\ 2013, \apjl, 779, L5 

\bibitem[Ortolani et al.(1993)]{ortolaniter1orig} Ortolani, S., Bica, E., \& Barbuy, B.\ 1993, \aap, 267, 66 

\bibitem[Ortolani et al.(1995)]{pal6vi} Ortolani, S., Bica, E., \& Barbuy, B.\ 1995, \aap, 296, 680 

\bibitem[Ortolani et al.(1997a)]{djorg2discov} Ortolani, S., Bica, E., \& Barbuy, B.\ 1997a, \aaps, 126, 319

\bibitem[Ortolani et al.(1997b)]{ortolanihp1vi} Ortolani, S., Bica, E., \& Barbuy, B.\ 1997b, \mnras, 284, 692 

\bibitem[Ortolani et al.(1997c)]{ortolaniter2} Ortolani, S., Bica, E., \& Barbuy, B.\ 1997c, \aap, 326, 614 

\bibitem[Ortolani et al.(1997d)]{ortolaniter4} Ortolani, S., Barbuy, B., \& Bica, E.\ 1997d, \aap, 319, 850 

\bibitem[Ortolani et al.(1998)]{ortolaniter12} Ortolani, S., Bica, E., \& Barbuy, B.\ 1998, \aaps, 127, 471 

\bibitem[Ortolani et al.(1999a)]{ortolaniter1} Ortolani, S., Barbuy, B., Bica, E., et al.\ 1999a, \aap, 350, 840

\bibitem[Ortolani et al.(1999b)]{ortolaniter9} Ortolani, S., Bica, E., \& Barbuy, B.\ 1999b, \aaps, 138, 267 

\bibitem[Ortolani et al.(2001a)]{nicmos1} Ortolani, S., Barbuy, B., Bica, E., et al.\ 2001a, \aap, 376, 878 

\bibitem[Ortolani et al.(2001b)]{ortolani6569} Ortolani, S., Bica, E., \& Barbuy, B.\ 2001b, \aap, 374, 564 

\bibitem[Ortolani et al.(2006)]{bh261discov} Ortolani, S., Bica, E., \& Barbuy, B.\ 2006, \apjl, 646, L115

\bibitem[Ortolani et al.(2007)]{nicmos2} Ortolani, S., Barbuy, B., Bica, E., Zoccali, M., \& Renzini, A.\ 2007, \aap, 470, 1043 

\bibitem[Ortolani et al.(2011)]{ortolaniao} Ortolani, S., Barbuy, B., Momany, Y., et al.\ 2011, \apj, 737, 31 

\bibitem[Ortolani et al.(2012)]{kron49} Ortolani, S., Bonatto, C., Bica, E., Barbuy, B., \& Saito, R.~K.\ 2012, \aj, 144, 147 

\bibitem[Paust et al.(2010)]{paustmf} Paust, N.~E.~Q., Reid, I.~N., Piotto, G., et al.\ 2010, \aj, 139, 476 

\bibitem[Pe{\~n}aloza et al.(2015)]{gc02spec} Pe{\~n}aloza, F., Pessev, P., Va{\'s}quez, S., et al.\ 2015, \pasp, 127, 329 

\bibitem[Piotto et al.(2002)]{piottosnap} Piotto, G., King, I.~R., Djorgovski, S.~G., et al.\ 2002, \aap, 391, 945 

\bibitem[Piotto et al.(2015)]{piottolegacy} Piotto, G., Milone, A.~P., Bedin, L.~R., et al.\ 2015, \aj, 149, 91 

\bibitem[Ross et al.(2014)]{holtziso} Ross, T.~L., Holtzman, J.~A., Anthony-Twarog, B.~J., et al.\ 2014, \aj, 147, 4 

\bibitem[Rossi et al.(2015)]{pmrossi} Rossi, L.~J., Ortolani, S., Barbuy, B., Bica, E., \& Bonfanti, A.\ 2015, \mnras, 450, 3270

\bibitem[Rutily \& Terzan(1977)]{origvars6638} Rutily, B., \& Terzan, A.\ 1977, \aaps, 30, 315 

\bibitem[Samra et al.(2012)]{samrapm} Samra, R.~S., Richer, H.~B., Heyl, J.~S., et al.\ 2012, \apjl, 751, L12

\bibitem[Saracino et al.(2015)]{liller1} Saracino, S., Dalessandro, E., Ferraro, F.~R., et al.\ 2015, \apj, 806, 152

\bibitem[Saracino et al.(2016)]{sara6624} Saracino, S., Dalessandro, E., Ferraro, F.~R., et al.\ 2016, \apj, 832, 48

\bibitem[Sarajedini et al.(2007)]{ataggc} Sarajedini, A., Bedin, L.~R., Chaboyer, B., et al.\ 2007, \aj, 133, 1658

\bibitem[Schlafly et al.(2018)]{decaps} Schlafly, E.~F., Green, G.~M., Lang, D., et al.\ 2018, \apjs, 234, 39

\bibitem[Siegel et al.(2011)]{acssgr} Siegel, M.~H., Majewski, S.~R., Law, D.~R., et al.\ 2011, \apj, 743, 20

\bibitem[Simunovic \& Puzia(2016)]{bssuv} Simunovic, M., \& Puzia, T.~H.\ 2016, \mnras, 462, 3401 

\bibitem[Smith \& Stryker(1986)]{rrlspec6638} Smith, H.~A., \& Stryker, L.~L.\ 1986, \pasp, 98, 453 

\bibitem[Stephens \& Frogel(2004)]{sf04} Stephens, A.~W., \& Frogel, J.~A.\ 2004, \aj, 127, 925 

\bibitem[Tang et al.(2017)]{baitian6553} Tang, B., Cohen, R.~E., Geisler, D., et al.\ 2017, \mnras, 465, 19

\bibitem[Trager et al.(1995)]{trager} Trager, S.~C., King, I.~R., \& Djorgovski, S.\ 1995, \aj, 109, 218 

\bibitem[Valencic et al.(2009)]{ter2x6} Valencic, L.~A., Smith, R.~K., Dwek, E., Graessle, D., \& Dame, T.~M.\ 2009, \apj, 692, 502 

\bibitem[Valenti et al.(2004)]{v04obs} Valenti, E., Ferraro, F.~R., \& Origlia, L.\ 2004, \mnras, 351, 1204 

\bibitem[Valenti et al.(2005)]{valenti05} Valenti, E., Origlia, L., \& Ferraro, F.~R.\ 2005, \mnras, 361, 272 

\bibitem[Valenti et al.(2010)]{v10} Valenti, E., Ferraro, F.~R., \& Origlia, L.\ 2010, \mnras, 402, 1729

\bibitem[Valenti et al.(2011)]{6569spec} Valenti, E., Origlia, L., \& Rich, R.~M.\ 2011, \mnras, 414, 2690 

\bibitem[Valenti et al.(2015)]{valentiter1} Valenti, E., Origlia, L., Mucciarelli, A., \& Rich, R.~M.\ 2015, \aap, 574, A80

\bibitem[VandenBerg et al.(2013)]{v13} VandenBerg, D.~A., Brogaard, K., Leaman, R., \& Casagrande, L.\ 2013, \apj, 775, 134 

\bibitem[Wagner-Kaiser et al.(2016)]{bayesianrachel} Wagner-Kaiser, R., Stenning, D.~C., Sarajedini, A., et al.\ 2016, \mnras, in press

\bibitem[Wagner-Kaiser et al.(2017)]{rachelage} Wagner-Kaiser, R., Sarajedini, A., von Hippel, T., et al.\ 2017, arXiv:1702.08856

\bibitem[Watkins et al.(2015a)]{hstpromo2} Watkins, L.~L., van der Marel, R.~P., Bellini, A., \& Anderson, J.\ 2015a, \apj, 803, 29 

\bibitem[Watkins et al.(2015b)]{hstpromo3} Watkins, L.~L., van der Marel, R.~P., Bellini, A., \& Anderson, J.\ 2015b, \apj, 812, 149

\bibitem[Wijnands et al.(2002)]{ter1x6} Wijnands, R., Heinke, C.~O., \& Grindlay, J.~E.\ 2002, \apj, 572, 1002 

\bibitem[Williams et al.(2014)]{phat} Williams, B.~F., Lang, D., Dalcanton, J.~J., et al.\ 2014, \apjs, 215, 9 

\bibitem[Zinn \& West(1984)]{ZW84} Zinn, R., \& West, M.~J.\ 1984, \apjs, 55, 45 

\bibitem[Zoccali et al.(2014)]{gibs} Zoccali, M., Gonzalez, O.~A., Vasquez, S., et al.\ 2014, \aap, 562, A66

\bibitem[Zoccali et al.(2016)]{gibs3} Zoccali, M., Vasquez, S., Gonzalez, O.~A., et al.\ 2016, arXiv:1610.09174

\end{thebibliography}
\end{document}